# Spin-Electric Control of Individual Molecules on Surfaces

Paul Greule[1], Wantong Huang[1], Máté Stark[1], Kwan Ho Au-Yeung[1,2], Johannes Schwenk[1], Jose Reina-Gálvez[3], Christoph Sürgers[1], Wolfgang Wernsdorfer[1,4], Christoph Wolf[5,6*], Philip Willke[1,2*]

[1] Physikalisches Institut (PHI), Karlsruhe Institute of Technology (KIT), Karlsruhe, Germany
[2] Center for Integrated Quantum Science and Technology (IQST), Karlsruhe Institute of Technology, Karlsruhe, Germany
[3] Department of Physics, University of Konstanz, Konstanz, Germany
[4] Institute for Quantum Materials and Technologies (IQMT), Karlsruhe, Germany
[5] Center for Quantum Nanoscience, Institute for Basic Science (IBS), Seoul, Republic of Korea.
[6] Ewha Womans University, Seoul, Republic of Korea.

* corresponding author: philip.willke@kit.edu, wolf.christoph@qns.science

**Individual magnetic molecules are promising building blocks for quantum technologies because of their chemical tunability, nanoscale dimensions, and ability to self-assemble into ordered arrays. However, harnessing their properties in quantum information processing requires precise local control of their spin properties. In this work, we present spin-electric coupling (SEC) for two molecular spin systems, iron phthalocyanine (FePc) and Fe-FePc complexes, adsorbed on a surface. We use electron spin resonance combined with scanning tunnelling microscopy (ESR-STM) to locally address them with the STM tip and electrically tune them using the applied bias voltage. These measurements reveal a pronounced nonlinear voltage dependence of the resonance frequency, linked to the energic onset of other molecular orbitals. We attribute this effect to a transport-mediated exchange field from the magnetic tip, providing a large, highly localized, and broadly applicable SEC mechanism. Finally, we demonstrate that the SEC enables all-electrical coherent spin control: In Rabi oscillation measurements of both single and coupled Fe-FePc complexes we show that the spin dynamics can be tuned, demonstrating a pathway towards electrically controlled quantum operation.**

Electron and nuclear spins in single molecules have attracted significant interest as potential building blocks for applications in spintronics, quantum sensing, and quantum computing[1]. Magnetic molecular systems are of nanoscopic size, benefit from self-assembly and offer unique structural as well as chemical tunability via modern synthetic chemistry[2,3]. A key challenge lies in achieving reliable and local control over individual spin centers. A potential solution is the use of electric fields, which – in contrast to magnetic fields – can be efficiently applied in a confined region. Thus, spin-electric coupling (SEC) in molecules has in recent years emerged as a promising control mechanism and was realized in a variety of systems[4–10]: These specifically tailored molecular platforms typically rely on structural distortions that modulate key parameters of the spin Hamiltonian, such as zero-field splitting, g-factor, hyperfine interaction or exchange coupling. Often, effective SEC requires a soft, electrically polarizable molecular environment and spin energy levels that are highly sensitive to structural changes. However, the experimentally observed shifts $\Delta f$ in spin resonance frequencies due to electric fields have so far remained relatively modest, amounting to less than $\frac{\Delta f}{f_0} < 1\%$ of the qubit's unperturbed resonance frequency $f_0$. Using a scanning tunneling microscope (STM), SEC has recently been demonstrated for individual atomic spins on surfaces[11] reaching up to $\frac{\Delta f}{f_0} \approx 3\%$ for individual Titanium atoms. Here, a combination of electron spin resonance (ESR) with STM has been used[12,13], which has emerged as a powerful tool providing sub-Ångström resolution while also permitting coherent control of single spin states[14–17]. However, the implementation of SEC in molecular systems for ESR-STM has been mostly addressed theoretically[18–21].

In this study, we use ESR-STM to investigate the SEC in two molecular spin systems on MgO/Ag(100) – iron phthalocyanine (FePc) and an Fe-FePc complex. We show that varying the DC bias voltage in the STM junction shifts the resonance frequency of the spin transition $f_{\text{res}} \propto V_{\text{DC}}$. In FePc, this shift becomes strongly non-linear – reaching close to 30% – when the electrochemical potential is close to the lowest unoccupied molecular orbital (LUMO). We attribute this behavior to a transport-mediated exchange interaction between the magnetic tip and the molecule spin[18–21], which predicts such a logarithmic divergence. This effect is not only large and highly localized (even more so than electric fields), but also universal and consequently broadly applicable to other molecular spin systems. In the second part, we demonstrate that the SEC can be readily employed in all-electrical coherent control of spin dynamics: Rabi oscillation measurements on Fe-FePc complexes reveal that individual spins can be selectively detuned using $V_{\text{DC}}$ only. Finally, we extend this to electric tuning of coupled dimer spins, highlighting the potential of SEC for individual control with nanoscale precision in larger spin assemblies.

**Results and Discussion**

An STM topography of the sample is shown in Figure 1a: We deposited Fe atoms and FePc molecules onto 2 monolayers of MgO atop an Ag(100) crystal. FePc molecules were shown to form a S=1/2 system that is localized on the central Fe atom[22]. Additionally, we include in this work a spin complex which consists of one FePc molecule that is strongly coupled to an adjacent Fe atom via one of its ligands. As shown previously[23], these Fe-FePc organometallic complexes can be built using tip-assisted assembly and form a mixed-spin (1,1/2) ferrimagnet with a well-separated doublet of $m_z = \pm\frac{1}{2}$, mimicking a S=1/2 system. Both FePc and Fe-FePc constitute ideal two-level systems that allow for coherent quantum control[15,23]. We employ both in this study, since the molecular orbital structure of FePc[22] demonstrates best the exchange bias mechanism, while the coherent control is facilitated on the Fe-FePc complex due to a resilience to inelastic electron scattering[23].

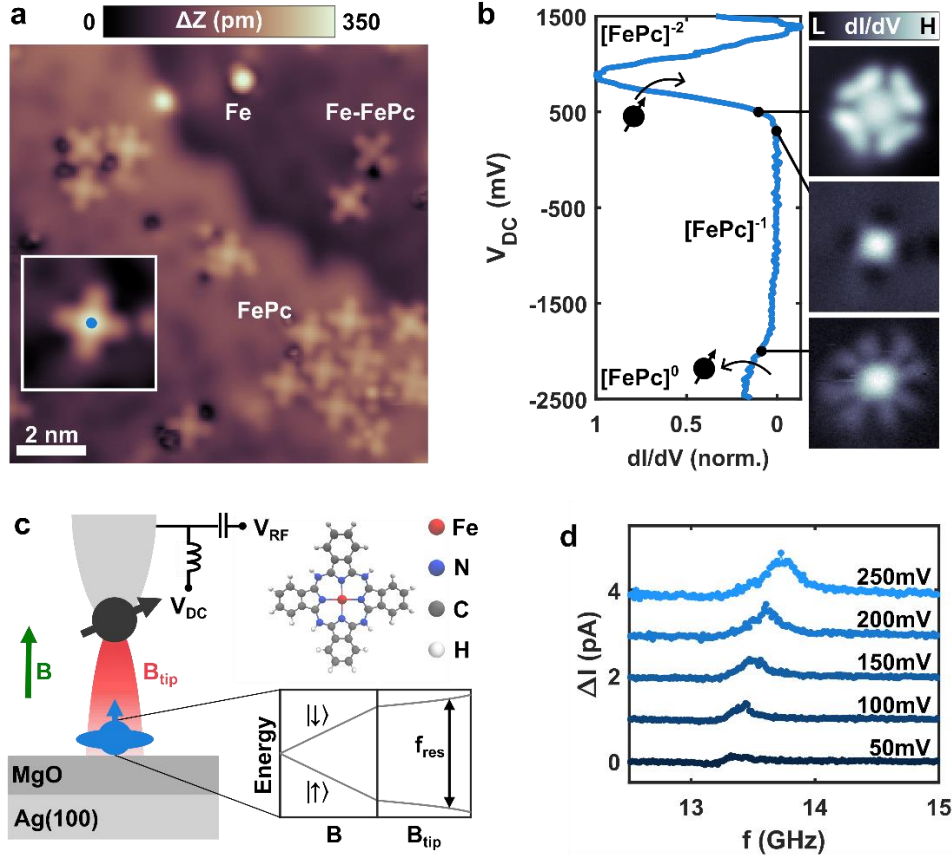

**Figure 1. Molecular spins on MgO/Ag(100). (a)** STM topography of the surface with the deposited Fe atoms and FePc molecules and a built Fe-FePc complex (Image conditions: $I = 20$ pA, $V_{DC} = -100$ mV). The inset shows a close-up topography of a single FePc molecule (2.2 nm × 2.2 nm, $I = 50$ pA, $V_{DC} = 100$ mV). The blue dot marks the tip position of the experiments shown in (b) and (d). **(b)** Left: Differential conductance (d$I$/d$V$) spectra acquired on the center of the FePc ($I_{set} = 5$ pA, $V_{set} = 100$ mV). The arrows indicate the removal (addition) of an electron leading to the transition from the [FePc]$^{-1}$ to [FePc]$^{0}$ and [FePc]$^{-2}$ charge state. The d$I$/d$V$ maps to the right show the spatial extent of both states ($V_{DC} = -2000$ mV and 500 mV) in comparison to the in-gap state (300 mV). **(c)** Left: Schematic drawing of the experimental setup with a spin-polarized STM tip above the FePc molecule atop MgO/Ag(100). A DC bias voltage $V_{DC}$ and radio-frequency (RF) voltage $V_{RF}$ are applied across the tunnel junction. The magnetic tip, realized by picking up individual Fe atoms from the surface, creates a highly localized tip-field $B_{tip}$ acting on the surface spin together with an externally applied magnetic field $B$. This is illustrated by the energy level diagram of the spin ½ on the right hand side. Top Right: Chemical configuration of the FePc molecule. **(d)** Electron spin resonance (ESR) measured on a FePc molecule recorded at different $V_{DC}$ showing the change in tunnel current $\Delta I$ as a function of frequency $f$ (ESR conditions: $I_{set} = 20$ pA, $V_{set} = 60$ mV, $B = 484$ mT, $V_{RF} = 10$ mV). The frequency sweeps were taken at constant height (open feedback loop) and are vertically shifted for clarity.

The electronic structure of the FePc molecule is characterized in Fig. 1b by differential conductance d$I$/d$V$ measurements (see also Ref.[22]). We observe pronounced conductance maxima around $-2000\ \mathrm{mV}$ and $700\ \mathrm{mV}$ related to the process of removing (adding) an electron to the molecule[24]. We assign these energy positions to the highest occupied molecular orbital (HOMO) and the lowest unoccupied molecular orbital (LUMO)[22,25], respectively (see also Supplementary Section 3). Density functional theory (DFT) calculations indicate that this electronic configuration involves one unpaired spin in the $a_{1g}$ orbital. This results from a charge transfer from the substrate ([FePc]$^{-1}$) and leads to a S=1/2 ground state[15,18,22] (see also Supplementary Section 4). The resulting magnetic spin state can be probed by ESR-STM (Fig. 1c). For ESR, the $m_z = \pm \frac{1}{2}$ ground states are split by an external magnetic field $B$ perpendicular to the sample surface

$$hf_{\mathrm{res}} = g\mu_{\mathrm{B}}(B + B_{\mathrm{tip}}) \qquad (1)$$

where $f_{\mathrm{res}}$ is the resonance frequency, $h$ is Planck's constant, $\mu_{\mathrm{B}}$ the Bohr magneton and $g \approx 2$ the g-factor (see Fig. 1c). Moreover, $B_{\mathrm{tip}}$ accounts for the influence of the highly localized magnetic tip field leading to a shift $\Delta f$ of the surface spin's resonance frequency. $B_{\mathrm{tip}}$ consists of both magnetic exchange and magnetic dipole-dipole interaction[26,27] which results in different amplitude and sign of $B_{\mathrm{tip}}$ depending on the particular magnetic tip apex.

Motivated by spin-electric coupling, which was recently observed for individual Ti atoms on MgO/Ag(100)[11] in ESR-STM, we investigated the dependence of the ESR signal as a function of bias voltage $V_{\mathrm{DC}}$ (Fig. 1d). Here, we keep the tip-sample distance and the external field $B$ constant while sweeping the ESR frequency. Indeed, we find a linear dependence $f_{\mathrm{res}} \propto V_{\mathrm{DC}}$ for the voltage range shown. This frequency shift can be interpreted as a contribution to the Zeeman energy by the SEC (sketch in Fig. 1c). The intensity of the ESR signal increases with $|V_{\mathrm{DC}}|$, mainly due to increasing tunneling current $I$ (see Supplementary Section 7 for discussion on the intensity of the ESR signal). For different magnetic tips, we observed that the linear voltage dependence occurs with a varying magnitude ranging from $0.5 - 8\ \mathrm{MHz/mV}$ (see Supplementary Section 6). In the case of Ti atoms, a linear shift of similar magnitude was explained by a piezoelectric coupling between the magnetic tip and the surface spin[11]: As a consequence of $V_{DC}$, the spin is displaced in the magnetic field of the tip which increases or decreases $B_{tip}$. This effect is additionally accompanied by a change in g-factor. In contrast, in recent works by some of the authors, it was shown that SEC can also result from transport-mediated exchange interaction[18,20,21], described in detail below. This was supported by first experimental data on Ti and FePc[19].

To elucidate the mechanism, we investigated the voltage dependence of the resonance frequency for FePc molecules in a large bias voltage range (Fig. 2), as well as for different magnetic tips (see also Supplementary Section 9). For the first magnetic tip in Fig. 2a, we find that the resonance peak position starts to shift drastically and in a non-linear manner at voltages above $\approx 250\,\text{mV}$. This shift $\Delta f$ is accompanied by an increase in peak linewidth and amplitude as well as a change in its asymmetry. Additionally, we find that $\Delta f$ changes sign when employing a tip of opposite magnetic field direction $B_{tip}$ (Fig. 2b). In Fig. 2c we compare the shift of the resonance frequency $\Delta f(V_{DC})$ for both datasets. Due to the non-linearity, the relative shift $\frac{\Delta f}{f_0}$ in Fig. 2c is rather large [$\pm(10-30)\%$] compared to previous works (~3% in Ref.[11] and ~0.2% in Ref.[5]). We stress that such non-linear behavior as found in Fig. 2 is not expected from a piezo-electric displacement model as previously used for Ti atoms[11]. Moreover, the latter relies on the electric-field induced displacement of the charged surface spin, and we find this to be incompatible with the observed sign of $\Delta f$ for FePc (see Supplementary Section 8).

In the following, we aim to explain the behavior by the indirect exchange interaction between the molecule and the magnetic tip, referred to as exchange bias[18,19,21,28–32]. Notably, the non-linear part of $\Delta f$ emerges when the applied $V_{DC}$ reaches the onset of the FePc LUMO shown in Fig. 1b. This onset hints towards a SEC mechanism which involves the influence of the unoccupied electronic states. A similar effect is found in quantum dot spin systems[28–30], for instance, in carbon nanotubes contacted with ferromagnetic electrodes[28]. This concept of exchange bias relies on virtual tunneling processes into the excited states and was recently described in the framework of ESR-STM[18–21]. Fig. 2d,e shows a schematic of the exchange bias in the tunneling junction for the two magnetic tip configurations. In both cases increasing $V_{DC}$ raises the electrochemical potential. Subsequently, the up (down) polarization of the spin-polarized tip enhances virtual tunneling of spin-up (down) electrons into the doubly occupied state (Fig. 2d,e, respectively). Due to the imbalance of the spin densities in the tip electrode, the virtual tunneling processes cause different energy corrections for the spin up and down state which adds to the Zeeman energy. This energy correction can be written as an effective exchange field[19,28]:

$$B_{\text{tip}} = -\frac{P\gamma_T}{2\pi}\ln\left(\left|\frac{\epsilon-eV_{DC}}{\epsilon+U-eV_{DC}}\right|\right) + B_0 \qquad (2)$$

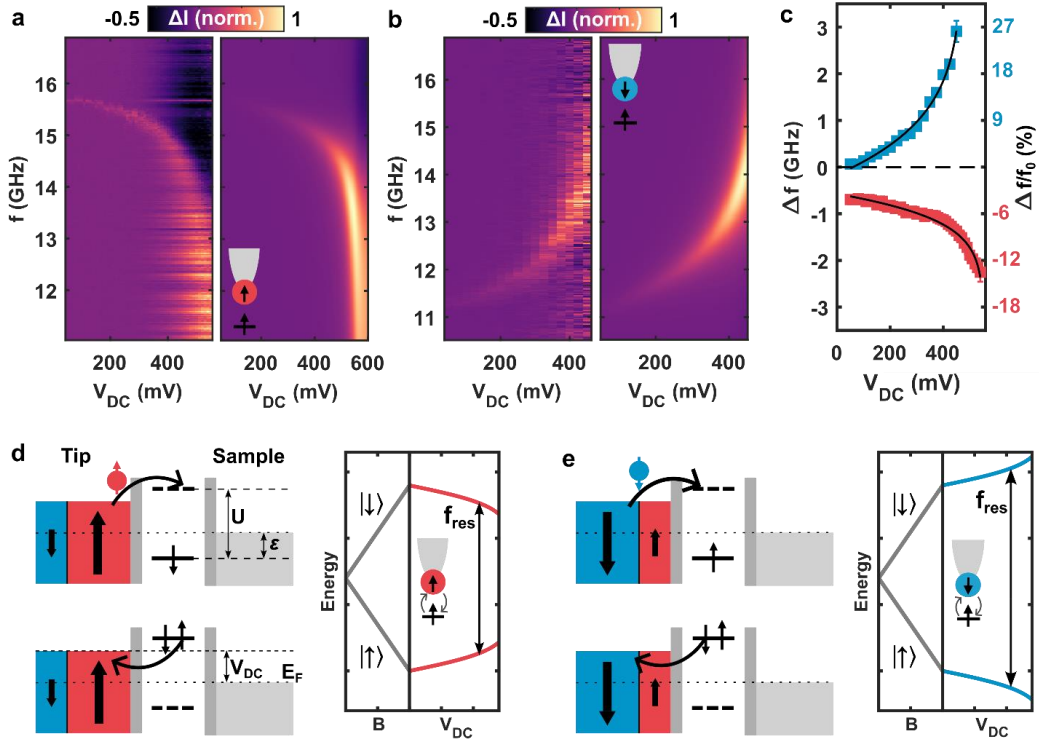

**Figure 2. Non-linear SEC in the ESR spectra on FePc. (a)** Colormap of the ESR signal $\Delta I$ as a function of $V_{DC}$ and $f$ on a single FePc molecule revealing a non-linear behavior at higher $V_{DC}$. The left panel shows the experimental data (ESR conditions: $I_{set} = 10$ pA, $V_{set} = 60$ mV, $B = 585$ mT, $V_{RF} = 8$ mV). The right panel displays a corresponding simulation according to the exchange bias model. The inset illustrates an STM tip with a spin polarization $P > 0$ used for the simulation. **(b)** ESR map $\Delta I(f, V_{DC})$ analogous to (a), but with a different magnetic tip with a spin polarization $P < 0$ ($I_{set} = 20$ pA, $V_{set} = 60$ mV, $B = 399$ mT, $V_{RF} = 8$ mV). **(c)** Frequency shift $\Delta f = f_{res} - 2\mu_B B/h$ over $V_{DC}$ extracted from the spectra in (a) (red) and (b) (blue). The black lines show corresponding fits of the exchange bias model, see Eq. (1) and (2). The second y-axis on the right hand side displays the relative change of the resonance frequency $\Delta f/f_0$ for the red and blue data set, respectively. **(d,e)** Schematic drawings of the virtual tunneling processes leading to the tip-induced exchange field: The molecular spin is described via a single impurity Anderson model (SIAM)[33,34]. The molecular energy levels, described by the ionization energy $\epsilon$ and the Coulomb repulsion energy $U$, lie between the electrochemical potential of the left spin-polarized tip electrode and the right sample electrode, separated by the vacuum tunneling barrier and the MgO layer, respectively. The bias voltage $V_{DC}$ moves the potential of the magnetic tip closer to the doubly occupied level for positive voltages. Subsequently, charge fluctuations and virtual tunneling processes (electron tunneling into the molecule and back) between the spin-polarized tip electrode and the molecule are enhanced. The polarization of the tip determines the dominating virtual tunneling (spin up in (d), spin down in (e)), that favor different spin states of the molecule. This leads to different $B_{tip}$ (see Supplementary Section 5 for details) as depicted by the energy level diagrams.

Here, the spin polarization $P = \frac{n_\uparrow - n_\downarrow}{n_\uparrow + n_\downarrow}$ quantifies the imbalance of the density of spin up ($n_\uparrow$) and down electrons ($n_\downarrow$) and sets the direction of the observed frequency shift ascribed to the tip. The coupling between the molecule and the tip $\gamma_T$ can be controlled in the experiment via the conductance setpoint $\gamma_T \propto G$ which alters the width of the vacuum barrier (see Supplementary Fig. S9). $e$ is the electron charge and $B_0$ accounts for a residual tip field, stemming for instance from magnetic dipole contributions. The logarithmic relation between $V_{DC}$ and the energy levels of the molecule, $\epsilon$ (ionization energy) and $\epsilon + U$ ($U$ is the Coulomb repulsion energy) results in a non-linear, diverging behavior close to these energy levels. Using this model, we can describe the experimental data in Fig. 2a (Fig. 2b) using a positive (negative) tip polarization: In Fig. 2c, we first use Eq. (1) and (2) to fit the non-linear divergence of $\Delta f(V_{DC})$. Here, we use the FePc HOMO level, obtained in Fig. 1b, and fix $\epsilon = -2000 \text{ meV}$. We subsequently find $(\epsilon + U) = 553 \text{ meV}$ and $477 \text{ meV}$ for positive ($P > 0$) and negative ($P < 0$) tip polarizations. This is in good agreement with the onset of the observed conductance maximum associated with the LUMO from Fig. 1b. We note that the transition from molecular orbitals to the single impurity Anderson model of the exchange bias model is not trivial. But due to the strong d-character of the spin-carrying orbital, which we examined in additional DFT calculations (see Supplementary Section 4), a single orbital model provides a good approximation. We reproduce the data in Fig. 2a,b by performing full transport simulations, which aim to capture all features of the ESR-STM spectrum. The results are additionally shown in Fig. 2a,b alongside the experimental data and show a close agreement in amplitude and resonance frequency of the peak (see Supplementary Section 9). The convincing match between experiment and theory in Fig. 2a-c supports that the SEC is a result of the exchange bias mechanism outlined above. We cannot exclude the presence of other contributions, for instance piezo-electric effects, that would add to a linear background. However, the strong non-linear divergence with close connection to the position of the LUMO, as well as the sign change for differently polarized tips, suggests that the exchange bias is the dominant mechanism.

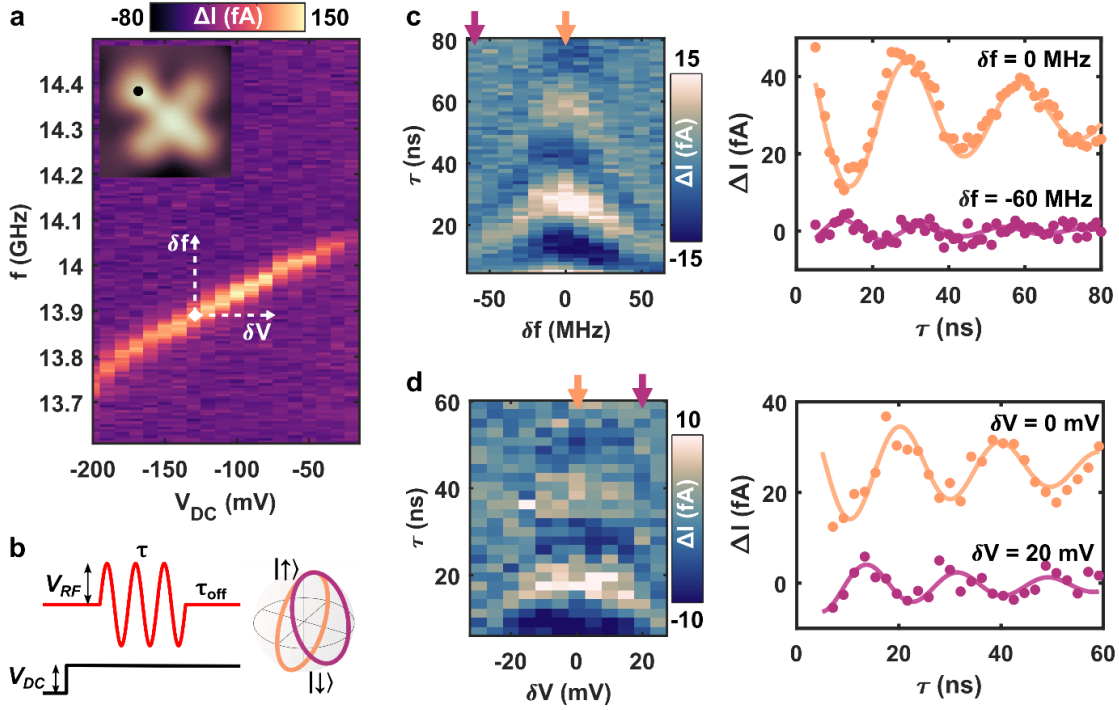

**Figure 3: Spin-electric Rabi detuning on a Fe-FePc complex. (a)** ESR colormap $\Delta I(f, V_{\rm DC})$ on the Fe-site of a Fe-FePc complex (ESR conditions: $I_{\rm set} = 6$ pA, $V_{\rm set} = -60$ mV, $B = 469$ mT, $V_{\rm RF} = 10$ mV). The tip position is marked by the black dot in the inserted topography. White arrows indicate the detuning in frequency $\delta f = f - f_{\rm res}$ and voltage $\delta V = V_{\rm DC} - V_{\rm set}$ from the resonance. **(b)** Left: Schematic drawing of the Rabi pulse scheme. The RF signal consists of an RF pulse with duration $\tau$ and amplitude $V_{\rm RF}$ followed by an off-time $\tau_{\rm off}$. The total cycle time $\tau_{\rm cycle} = \tau + \tau_{\rm off}$ is kept constant and a DC voltage $V_{\rm DC}$ is applied continuously for readout. Right: Bloch sphere representation of the spin evolution on resonance (orange) and off resonance (purple) **(c)** Rabi oscillations for different frequency detuning $\delta f$ (Rabi conditions: $I_{\rm set} = 4$ pA, $V_{\rm set} = -60$ mV, $B = 473$ mT, $V_{\rm RF} = 60$ mV, $f = 14.04$ GHz, $\tau_{\rm cycle} = 250$ ns). Left: Colormap of $\Delta I$ as a function of $\delta f$ and $\tau$ revealing the characteristic chevron pattern. The arrows refer to the traces shown on the right. Right: Single traces on (orange) and off (purple) resonance, plotting $\Delta I$ as a function of $\tau$. Solid lines are fits based on Eq. (3). Traces are vertically shifted for clarity. **(d)** Rabi oscillations for detuning the voltage $\delta V$ instead of $\delta f$ (Rabi conditions: $I_{\rm set} = 5$ pA, $V_{\rm set} = -60$ mV, $B = 450$ mT, $V_{\rm RF} = 20$ mV, $f = 14.25$ GHz, $\tau_{\rm cycle} = 400$ ns). Left: Colormap of $\Delta I(\delta V, \tau)$ showing a distorted chevron pattern. Right: Single traces analogous to (c). We note that the datasets presented in (a), (c) and (d) were measured separately with different tips on different molecules.

To further demonstrate that a strong SEC enables all-electrical spin control, we utilize the SEC in coherent control schemes, for which we employ Fe–FePc complexes (Fig. 1a and Fig. 3a inset). The ESR colormap in Fig. 3a shows the shift of $f_{\text{res}}$ as a function of $V_{\text{DC}}$. For the complex we also find an onset of non-linear behavior (see Supplementary Section 10) while performing ESR at $|V_{\text{DC}}| > 300 \text{ mV}$ remains challenging. Nevertheless, the spin complex is generally easier to use in coherent control experiments than pristine FePc. The pulse scheme used for Rabi oscillation measurements is depicted in Fig. 3b[14,15]. The resulting coherent oscillation of the spin state leads to a change in tunnel current $\Delta I$ as a function of the RF pulse duration $\tau$[35]:

$$\Delta I = A \cdot \sin(\Omega \cdot \tau + \phi) \cdot e^{-\tau/T_2} \tag{3}$$

With the amplitude $A$, the Rabi rate $\Omega$, the Rabi phase $\phi$ and phase coherence time $T_2$. Moreover, detuning from resonance $\delta f = f - f_{\text{res}}$ leads to a change in both amplitude $A$ and Rabi rate $\Omega$ of the observed oscillation:

$$\Omega = \sqrt{\Omega_0^2 + \delta f^2} \quad , \quad A = A_0 \frac{\Omega_0^2}{\Omega^2} \tag{4}$$

Here $A_0$ and $\Omega_0$ are the parameters at $f_{\text{res}}$. Consequently, the Rabi oscillations $\Delta I(\tau)$ can be tuned by $\delta f$ resulting in the typical chevron pattern (Fig. 3c). Utilizing the SEC, we now realize an all-electrical detuning via a change in voltage $\delta V = V_{\text{DC}} - V_{\text{set}}$ (Fig. 3d) while keeping $\delta f = 0$. We obtain a chevron pattern as well for $\Delta I(\tau, \delta V)$, in which the amplitude $A$ (Rabi rate $\Omega$) decreases (increases) for increasing $|\delta V|$. Compared to the frequency tuning, the pattern is slightly distorted. We attribute this to a linear contribution to $\Omega_0 \propto V_{DC}$ predicted for spin resonance in the exchange bias model[18,21]. Additionally, we expect a dependence of the amplitude with tunneling current $A \propto I \propto V_{DC}$ (see Supplementary Section 11 for details).

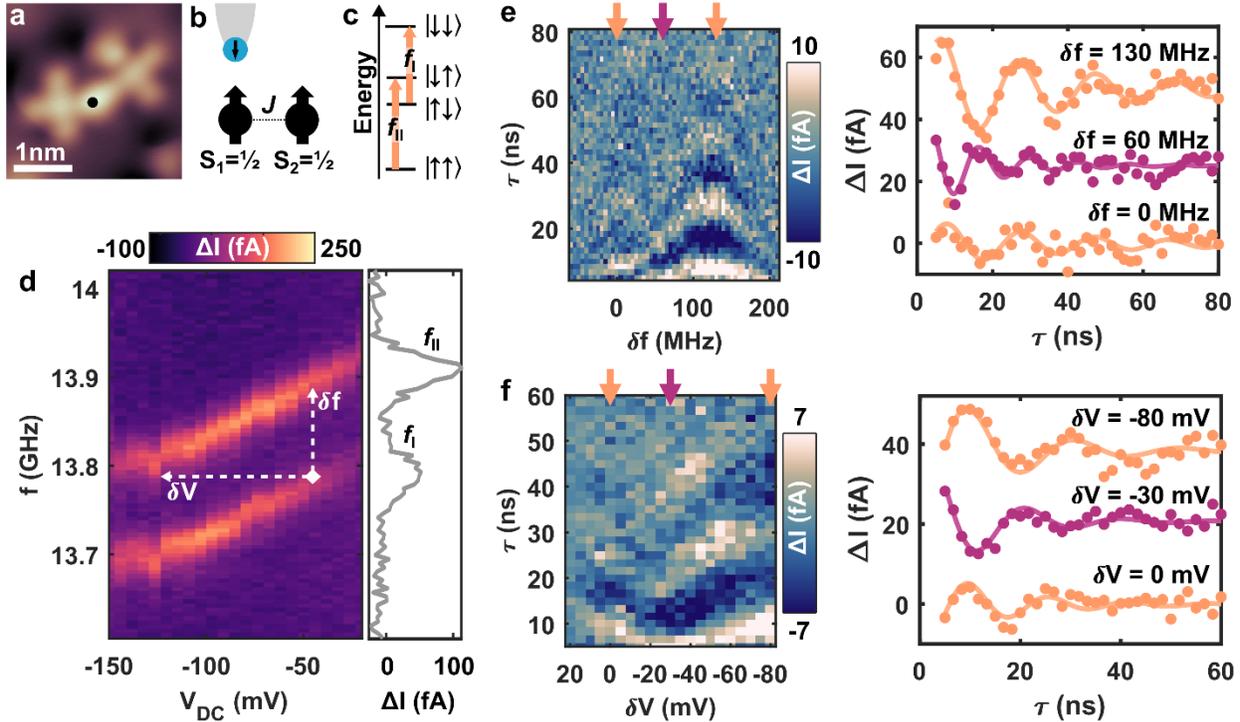

**Figure 4: Spin-electric Rabi detuning in a coupled spin system. (a)** Topography of two coupled Fe-FePc complexes (Image conditions: $I = 10 \text{ pA}, V_{DC} = -100 \text{ mV}$). The black dot marks the tip position of the subsequent measurements. **(b)** Schematic drawing of the two spin ½ with their exchange coupling $J$ and the tip above the first spin $S_1$. **(c)** Schematic energy level diagram of the combined spin states with the two ESR transitions $f_I$ and $f_{II}$. **(d)** ESR colormap $\Delta I(f, V_{DC})$ measured on the coupled spin system (ESR conditions: $I_{set} = 8 \text{ pA}, V_{set} = -40 \text{ mV}, B = 462 \text{ mT}, V_{RF} = 10 \text{ mV}$). A single frequency sweep (right panel) at $-50 \text{ mV}$ reveals two distinct peaks corresponding to $f_I$ and $f_{II}$, i.e. transitions corresponding to different spin states of the remote spin $S_2$. White arrows indicate the detuning in frequency $\delta f$ and voltage $\delta V$. The energy splitting of ~130 MHz remains constant across the whole voltage range, indicating that the spin-spin coupling ($S_1 - S_2$) is unaffected by the SEC. **(e)** Frequency detuning of Rabi oscillations (Rabi conditions: $I_{set} = 8 \text{ pA}, V_{set} = -40 \text{ mV}, B = 458 \text{ mT}, V_{RF} = 80 \text{ mV}, f = 13.97 \text{ GHz}$). Left: Colormap of $\Delta I$ as a function of $\delta f$ and $\tau$. The pattern shows two chevrons corresponding to the two ESR transitions. The weaker intensity of the left chevron arises from the low thermal population of the excited state $|\downarrow\rangle$ of $S_2$. The arrows at the top refer to the single traces to the right. Right: Single traces of $\Delta I$ versus $\tau$ for three $\delta f$. Solid lines show

fits to the circular datapoints based on Eq. (3). The traces were shifted vertically for clarity. **(f)** Electric detuning of Rabi oscillations, analogous to (e) (Rabi conditions: $I_{set} = 8 \text{ pA}, V_{set} = -40 \text{ mV}, B = 462 \text{ mT}, V_{RF} = 70 \text{ mV}, f = 14.04 \text{ GHz}$). Left: Colormap of $\Delta I(\delta V, \tau)$ showing the continuous electrical tuning from one ESR transition to the other. Right: $\Delta I$ as a function of $\tau$ for three different $\delta V$.

Finally, we realize the SEC detuning in a two-spin system: In Fig. 4a, two complexes are brought into proximity using tip-assisted manipulation to establish a coupled spin system. The resulting configuration (Fig. 4b) consists of a readout spin $S_1$ that is ferromagnetically coupled to the second spin $S_2$ mainly through Heisenberg exchange interaction $J$ (see Supplementary Section 12). Since the coupling is significantly smaller than the Zeeman energy, the system exhibits four distinct energy levels (Fig. 4c). The two resulting ESR transitions $f_I$ and $f_{II}$ (Fig. 4c,d) primarily reflect the alignments of $S_2$ in either $|\uparrow\rangle$ and $|\downarrow\rangle$ state[36,37]. $f_I$ and $f_{II}$ shift again as a function of $V_{DC}$ with the exchange bias from the tip acting on $S_1$. In the corresponding Rabi oscillation measurements (Fig. 4e) we now tune from $f_I$ to $f_{II}$ by changing $\delta f$ which leads to two chevron patterns. Again, the SEC enables all-electrical detuning via a change in voltage $\delta V$ (Fig. 4f). The main limitation in this approach is the increased tunneling current at higher voltages, which induces spin relaxation and decoherence[14,15]. However, the bias-controlled exchange field still permits to tune from the first transition ($\delta V = -20 \text{ mV}$) to the second ($\delta V = -80 \text{ mV}$).

**Conclusion**

Our measurements highlight that molecular spin systems can be tuned electrically via the bias voltage $V_{DC}$. In particular, the ESR measurements near the LUMO of FePc suggest that the strong non-linear SEC arises from the exchange bias due to enhanced virtual tunneling. While this does not exclude the existence of other contributions, our results indicate that in the present molecular systems, exchange bias is the dominant effect. Notably, the exchange bias mechanism has several important implications: First, unlike piezoelectric models, it does not require displacement of the molecule or one of its components, which extends it to a broader class of molecular systems, but also e.g. rigid solid-state defects. Second, it permits the integration of molecular spins into devices, where they can be readily tuned via the polarization of nearby ferromagnetic electrodes. Third, the SEC strength observed here reaching close to ~30% is significantly larger than most reported electric tuning effects for molecular spins. We believe that this can be further improved by synthesizing molecules with tailored molecular orbital structure, e.g. states lying close to the Fermi level. Finally, the results in Fig. 3 and 4 demonstrate not only the feasibility of combining coherent spin control with SEC, but also the ability to electrically tune one spin relative to another with nanometer precision. This precision arises from the particularities of the exchange bias: While a pure electric field from the tip apex would still act over distances exceeding tens of nanometers, the interplay between exchange interaction and the bias voltage – mediated by the magnetic tip electrode – localizes the SEC to the sub-nanometer scale. Further analysis of the data in Fig. 4 shows that only the molecular spin under the tip is tuned, while the other one stays completely unaffected (see Supplementary Section 12). Crucially, the ability to tune between two distinct spin resonances represents a key step toward conditional spin control in coupled spin systems. Consequently, the potential for tuning spin dynamics via the exchange bias paves the way for fast all-electrical gate operations in larger molecular quantum systems.

**Associated Content**

Supporting Information: Sample Preparation, Methods and Data Evaluation, d$I$/d$V$ Measurements on FePc and Fe-FePc, Density Functional Theory (DFT) of FePc, The Exchange Bias Model, Linear Spin Electric Coupling (SEC) of FePc and Fe-FePc, Polarity Preference of FePc and Fe-FePc, Piezo-Electric Displacement, Non-linear Spin Electric Coupling of FePc, Non-linear Spin electric coupling of Fe-FePc, Rabi Detuning Measurements, SEC in a coupled spin system, Rabi measurements in a coupled spin system

**Acknowledgments**

P.W. acknowledges funding from the Emmy Noether Programme of the DFG (WI5486/1-2) and financing from the Baden Württemberg Foundation Program on Quantum Technologies (Project AModiQuS). P.G. and P.W. acknowledge financial support from the Hector Fellow Academy (Grant No. 700001123). C.W. acknowledge support from the Institute for Basic Science (IBS-R027-D1). P.W. and K.H.A.Y acknowledge support from the Integrated Center for Quantum Science and Technology (IQST). P.W. and J.S. acknowledge funding from the ERC Starting Grant ATOMQUANT.

**Author contributions.** P.G. and P.W. conceived the experiment. P.G., W.H., M.S., K.H.A.Y., J.S., C.S., W.W. and P.W. set up the experiment and conducted the measurements. P.G., W.H., M.S., K.H.A.Y. and P.W. analyzed the experimental data. C.W. and J.R-G. developed the exchange bias model and made the simulations with P.G. C.W. performed the DFT calculations. P.G. and P.W. wrote the manuscript with input from all authors. W.W. and P.W. supervised the project.

**Notes.**

The authors declare no competing financial interests.

# Supplementary Information
# Spin-Electric Control of Individual Molecules on Surfaces

Paul Greule[1], Wantong Huang[1], Máté Stark[1], Kwan Ho Au-Yeung[1,2], Johannes Schwenk[1], Jose Reina-Gálvez[3], Christoph Sürgers[1], Wolfgang Wernsdorfer[1,4], Christoph Wolf[5,6*], Philip Willke[1,2*]

[1] Physikalisches Institut (PHI), Karlsruhe Institute of Technology (KIT), Karlsruhe, Germany

[2] Center for Integrated Quantum Science and Technology (IQST), Karlsruhe Institute of Technology, Karlsruhe, Germany

[3] Department of Physics, University of Konstanz, Konstanz, Germany

[4] Institute for Quantum Materials and Technologies (IQMT), Karlsruhe, Germany

[5] Center for Quantum Nanoscience, Institute for Basic Science (IBS), Seoul, Republic of Korea.

[6] Ewha Womans University, Seoul, Republic of Korea.

\* corresponding author: philip.willke@kit.edu, wolf.christoph@qns.science

**Table of Contents**

## 1. Sample Preparation

Sample preparation and all experiments were carried out in a Unisoku USM1600 system with a home-built dilution refrigerator. The measurements shown in Fig. 1 and 2 were measured at a base temperature of $1$ K, while the measurements of Fig. 3 and 4 were measured at $50$ mK. The in-situ sample preparation was performed under ultra-high vacuum conditions with a base pressure of $< 5 \times 10^{-10}$ mbar. The Ag(100) single crystal was first cleaned through multiple cycles of argon ion sputtering and subsequent annealing using an electron beam. MgO was grown by evaporating Mg in an oxygen-rich atmosphere ($\sim 1 \times 10^{-6}$ mbar) while maintaining the substrate at $510°$ C. A deposition time of $10$ min resulted in partial coverage ($\sim 50\%$) with MgO islands ranging from 2 to 5 monolayers in thickness. FePc molecules were deposited onto the surface using a home-built Knudsen cell (deposition time: $90$ s, pressure: $9 \times 10^{-10}$ mbar). Afterwards, Fe atoms were deposited onto the cooled sample by electron-beam evaporation for $21$ s.

## 2. Methods and Data Evaluation

For the ESR measurements and Rabi measurements we prepared spin-polarized tips by picking up $1 - 20$ Fe atoms from the MgO surface. In most cases the ESR active tips also showed strong asymmetries in d$I$/d$V$ spectra around 0 V on FePc molecules. For the experiments shown, we applied the DC and RF voltage to the tip and corrected this to the convention that the voltage is applied to the sample. The RF signal was generated by a Rhode & Schwarz SMB100B generator and mixed with the DC bias voltage $V_{\text{DC}}$ via a Marki Microwave MDPX-0305 Diplexer. For ESR, a magnetic field $B$ is applied perpendicular to the sample surface. The presented ESR frequency sweeps were measured with an on/off modulation at $323$ Hz where the RF voltage $V_{\text{RF}}$ in continuous-wave (cw) mode was only present in the A-cycle. The applied $V_{\text{DC}}$ was present in A- and B-cycle and therefore applied during the whole measurement. During the voltage dependent frequency sweeps the feedback of the STM controller was turned off to keep the position of the tip the same when altering $V_{\text{DC}}$. The signal was read out via a Stanford Research Systems SR860 digital lock-in amplifier. To analyze the frequency sweeps of our experiments we fitted the following Fano-function to our resonance peaks:

$$\Delta I = \frac{A}{q^2+1} \frac{(q \cdot \epsilon + 1)^2}{1+\epsilon^2} + c \qquad \text{with} \qquad \epsilon = \frac{f-f_{\text{res}}}{0.5 \cdot \Gamma} \qquad (S1)$$

with the amplitude $A$, the resonance frequency $f_{\text{res}}$, the linewidth (FWHM) $\Gamma$ and the $q$-factor which captures the asymmetry of the resonance peak.

For the Rabi measurements presented in Fig. 3 and 4 we followed the detection scheme introduced by Ref.[1]: Instead of a cw RF signal, we applied pulse trains (see Fig. 3b) with on-time $\tau$ and off-time $\tau_{\text{off}}$ keeping the sum ($\tau_{\text{cylce}}$) fixed. The RF pulses were only present in the A-cycle and triggered with a Zurich Instruments HDAWG. Therefore, the measured signal presents an average of the spin state-related tunnel current. For each data point $\tau$, the signal was averaged for $3\,\mathrm{s}$. After the data acquisition, a linear background[1] caused by current rectification with increasing pulse duration was subtracted.

### 3. d$I$/d$V$ Measurements on FePc and Fe-FePc

Here, we provide additional information on the electronic structure of the two probed spin systems. Fig. S1a shows the d$I$/d$V$ signal of FePc and the Fe-FePc complex. As mentioned in the main text, for FePc the conductance peaks are located around $-2000\,\mathrm{mV}$ and $700\,\mathrm{mV}$, respectively.

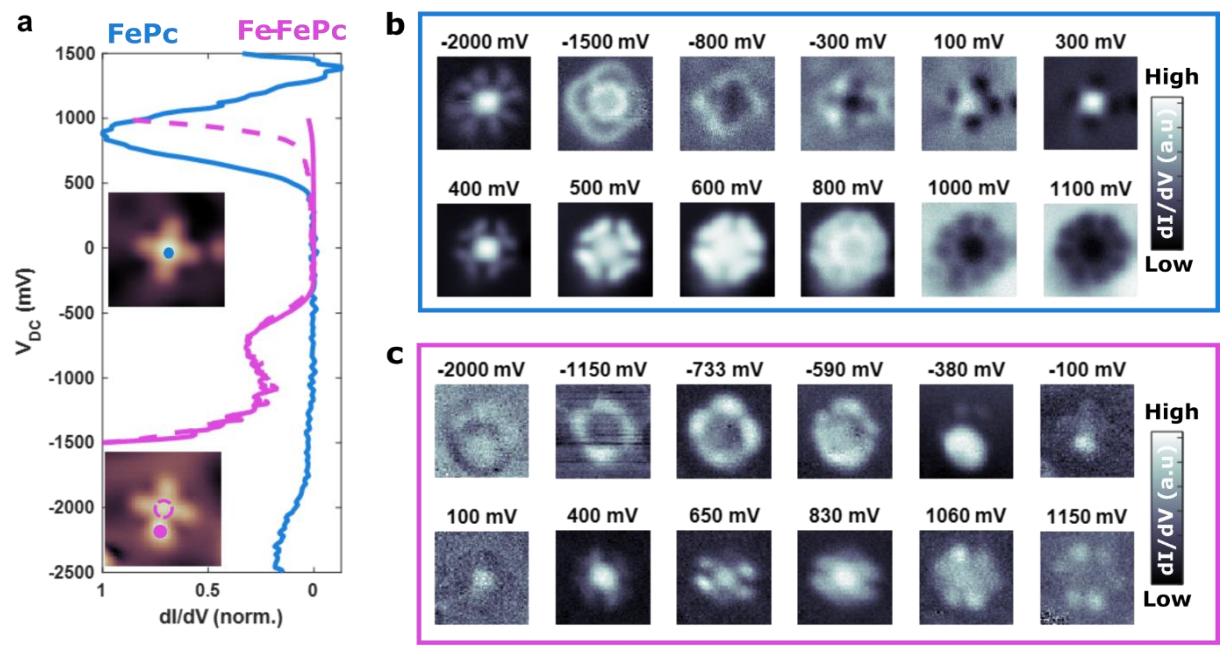

**Supplementary Fig. S1.** (a) Normalized differential conductance (d$I$/d$V$) spectra acquired on an isolated FePc molecule (blue) and a Fe–FePc complex (magenta) as the numerical derivative of the $V_{\mathrm{DC}}$ dependent tunnel current $I$. The insets show corresponding STM topographies with colored dots indicating the tip positions of the spectra. (b) Spatially resolved d$I$/d$V$ maps of the single FePc molecule shown in (a). The signal is recorded in closed loop at the bias voltages $V_{\mathrm{DC}}$ via a lock-in modulation $V_{\mathrm{mod}}$ (2.2 nm × 2.2 nm, $I_{\mathrm{set}} = 50\,\mathrm{pA}$, $V_{\mathrm{mod}} = 10\,\mathrm{mV}$). (c) d$I$/d$V$ maps of different $V_{\mathrm{DC}}$ of the Fe–FePc complex shown in (a) (2.5 nm × 2.5 nm, $I_{\mathrm{set}} = 5\,\mathrm{pA}$, $V_{\mathrm{mod}} = 10\,\mathrm{mV}$).

In contrast, we find for the Fe-FePc complex the HOMO closer to the Fermi level at $-600\,\mathrm{mV}$, while the LUMO is $> 1000\,\mathrm{mV}$. To obtain spatial information on the conductance, we performed d$I$/d$V$ maps, visible in Fig. S1b and c. For FePc we find the signal at $-2000\,\mathrm{mV}$ located at the center of the molecule, reflecting the HOMO. On the positive side, we see that with increasing voltage the signal spreads from the center towards the lobes, suggesting that tunneling into different orbitals contribute to the pronounced conductance peak[2]. For Fe-FePc we obtain a similar emerging behavior for the maximum at negative bias located on the Fe adatom of the complex. In contrast, the signal at positive bias is dominant on the FePc site of the complex. We note that the association of the spatial shape of the d$I$/d$V$ signal to molecular orbitals is not straightforward[3], since the pronounced signals observed in d$I$/d$V$ measurements correspond to charge transitions which are best described via Dyson orbitals and may differ from the Kohn-Sham spin orbitals.

## 4. Density Functional Theory (DFT) of FePc

We use DFT to correlate the charge and spin states of the FePc molecule with the charge states of a single impurity Anderson model (SIAM Fig. S2a). When adsorbed on 2 ML of MgO, the ground state of FePc has been found to be singly negatively charged, i.e. $[FePc]^{-1}$ and $S=1/2^2$. This corresponds to the ground state of the SIAM. To understand the charge transition points (i.e. $\varepsilon$ and $\varepsilon + U$) we perform charge-constrained SCF calculations, often also referred to as ΔSCF calculations[4]. Calculating the DFT total energy of the positively charged $[FePc]^0$ (S=1) and doubly negatively charged $[FePc]^{-2}$ (S=0) states allows to estimate the ionization potential (IP) and the electron affinity (EA) of the system. Their difference is the charging energy $U = 2.6\,eV$ (see Fig. S2b). We note that this value depends on several DFT parameters such as the van der Waals correction or the value of Hubbard-U.

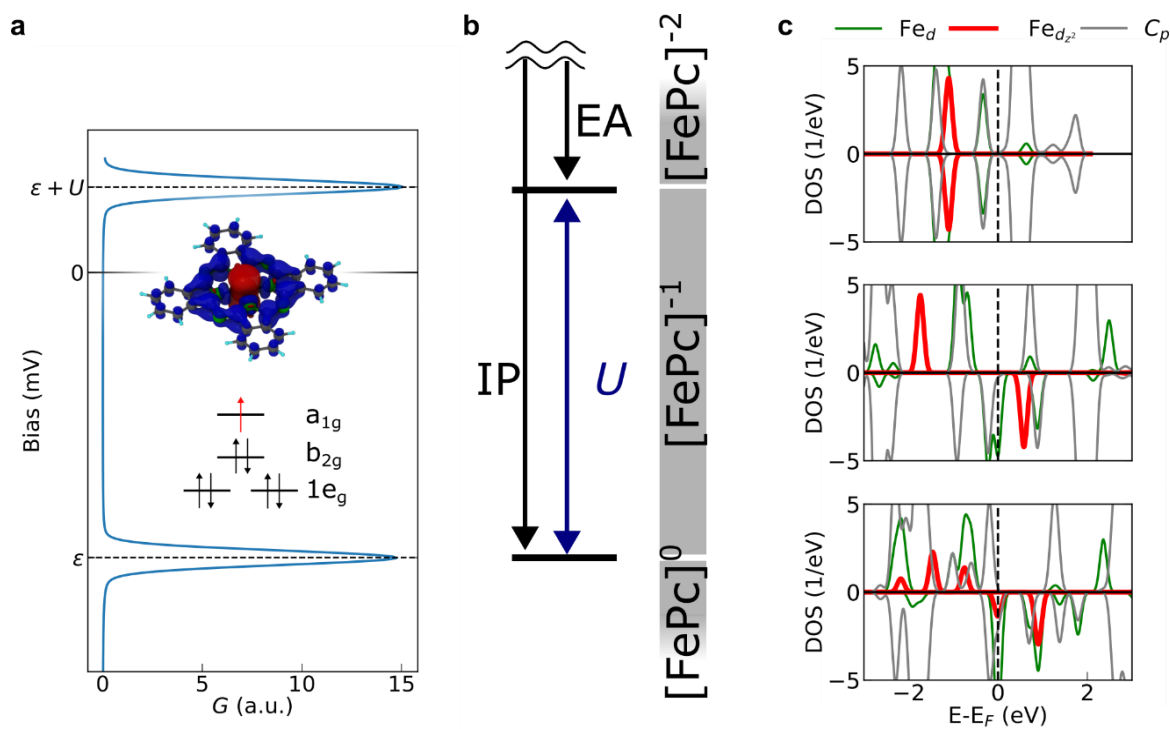

**Supplementary Fig. S2. (a)** Schematic representation of the spectral function for a single impurity Anderson model (SIAM) with the two relevant charge transitions at ionization energy $\varepsilon$ and charging energy $\varepsilon + U$. The inset shows the energy level diagram of the $[FePc]^{-1}$ charge state with the single occupied $a_{1g}$ orbital which has a strong $d_{z^2}$ contribution (71%). The spin polarization iso-surface above demonstrates the strong $d_{z^2}$ character of the orbital **(b)** Correspondence to the DFT ΔSCF calculations, indicating ionization potential (IP) and electron affinity (EA); their difference is $U$. **(c)** Calculated spin-polarized projected DOS (PDOS for the three charge configurations: $[FePc]^{-2}$, $[FePc]^{-1}$ and $[FePc]^0$.

Interestingly, the change in orbital occupation in these charge transition points indicates the addition or removal of an electron from an orbital with strong (~71%) $d_{z^2}$ character, that is predominantly localized at the Fe center. This is illustrated in the

PDOS in Fig. S2c and justifies, to a good approximation, the mapping of the charge and spin transitions of the [FePc]$^{-1}$ molecule with a$_{1g}$ character to a single d-like orbital, as it is often considered in the SIAM.

Details of the DFT calculation: DFT calculations were performed with an FePc molecule placed in a 20x20x20 Angstrom vacuum box. We used plane-wave DFT with pseudopotentials as implemented in Quantum Espresso. Pseudopotentials from the PSL library were used for all atoms and the basis was expanded with a cutoff of 90 (900) Ry for the kinetic energy (charge density). The PBE generalized gradient approximation was used to tread exchange and correlation. We used rVV10 for dispersive forces and Markov-Payne correction to decouple the cells from their periodic image. The relaxed geometry of the singly charged [FePc]$^{-1}$ was used for the ΔSCF calculations and no further relaxations were taken into account (sudden approximation for the charge transitions). The spin ground state was obtained from constrained DFT calculations of all possible spin states. The Brillouin zone was sampled using only the Γ-point and no Hubbard-U correction on the Fe was used in this calculation.

## 5. The Exchange Bias Model

As discussed in the main text, the theory of the exchange bias model that we use to discuss our data, has been originally established in the framework of quantum dots[5–9], but in recent theoretical works by some of the authors adapted for the framework of STM and single atoms and molecules[10–12]. As discussed in Supplementary Section 4, spins in these systems can be well described by a single impurity Anderson model (SIAM)[13,14]. As illustrated by the density of states in Fig. S3a, the single occupied orbital lies below the Fermi Energy $E_\mathrm{F}$ with the ionization energy $\epsilon$. Since the singly charged state is the ground state, $\epsilon$ needs to be negative, $\epsilon < 0$. The doubly occupied orbital appears at positive energy. Here, the energy difference between the orbitals is given by the Coulomb repulsion energy $U$. In our model, the chemical potential of the molecule aligns with the substrate electrode due to the adsorption on the MgO surface. The SIAM is used to describe the virtual particle exchange between the impurity and the spin baths from the tip and surface. Fig. S3b shows a schematic drawing of the tunneling junction. The energy levels from the SIAM are situated between the spin-polarized tip electrode (left) and the non-magnetic sample electrode (right), separated by two tunnelling barriers—vacuum on the tip side and a thin MgO layer on the sample side. We consider a virtual tunneling process between the tip electrode and the impurity. During this process, an electron tunnels into or out of the impurity and then back. Because the tip is spin-polarized, the probabilities for these virtual transitions differ for spin-up and spin-down state of the molecule. By applying a bias voltage $V_\mathrm{DC}$ we shift the chemical potential of the tip relative to the molecule, thereby tuning the rates of virtual electron addition (positive bias) or removal (negative bias). This results in spin-dependent energy corrections to the molecule's spin states, modifying the effective energy splitting beyond the contribution of the external magnetic field $B$.

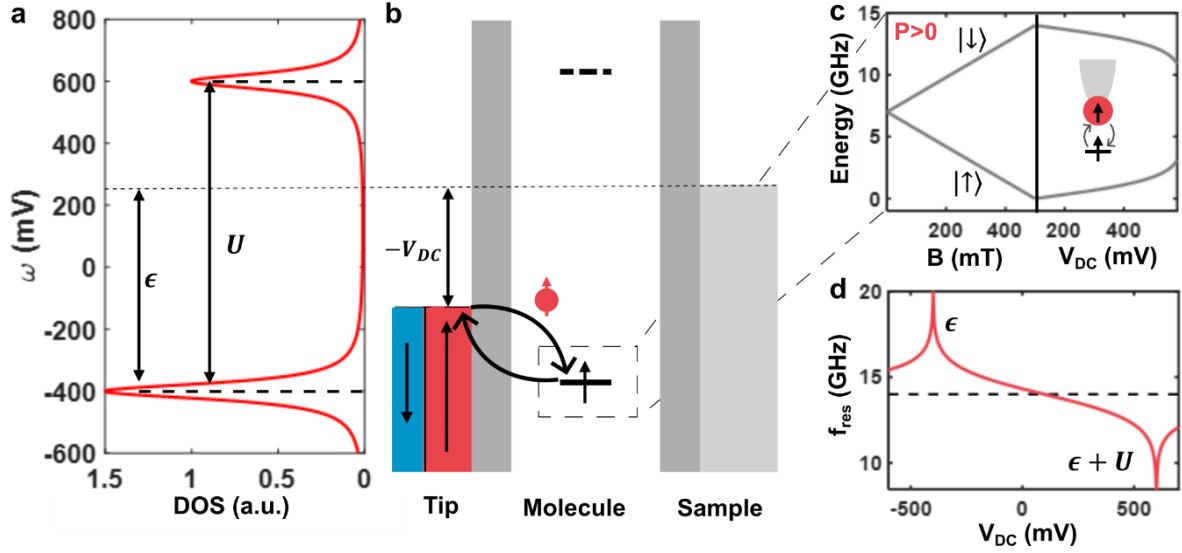

**Supplementary Fig. S3. (a)** Density of states (DOS) of the single impurity Anderson model (SIAM) with the ionization energy $\epsilon$ and the Coulomb repulsion energy $U$. **(b)** Schematic drawing of the tunnel junction. The energy levels from the SIAM are located between the spin-polarized tip (left) and the non-magnetic sample electrode (right). The energy barriers stemming from the vacuum and MgO are illustrated by dark grey boxes. The applied bias voltage $V_{DC}$ shifts the tip's chemical potential relative to the molecule and sample. The virtual tunneling process which mediates the exchange coupling is illustrated by the black curved arrows. **(c)** Energy level diagram of a spin under the influence of an external magnetic field $B$ as well as $V_{DC}$ for a positively spin-polarized tip ($P > 0$). The Zeeman splitting caused by $B$ is followed by an energy correction due to the transport mediated exchange coupling as a function of $V_{DC}$. **(d)** Resonance frequency $f_{res}$ calculated for a spin ½ system within the SIAM as a function of $V_{DC}$ ($g = 2, B = 0.5\,\text{T}, P = 1, \gamma_T = 0.36\,\text{T}, \epsilon = -400\,\text{meV}, \epsilon + U = 1000\,\text{meV}, B_0 = 0\,\text{T}$). As a reference, the dashed black line shows the resonance frequency in the absence of tip-field. The exchange bias features a logarithmic divergence at $\epsilon$ and $\epsilon + U$.

An exemplary energy level diagram for $P > 0$ is shown in Fig. S3c. The corrections can be described by an effective exchange field as presented in Eq. (2) in the main text. As shown in Fig. S3d, this leads to a linear shift of the resonance frequency $f_{res}$ with small bias voltage $V_{DC}$ and logarithmic divergences near the impurity energy levels $\epsilon$ and $\epsilon + U$. We note that the formula in Eq. (2) is only valid for $B > 0$ and we use the convention of the spin up state as ground state.

## 6. Linear Spin Electric Coupling (SEC) of FePc and Fe-FePc

We investigated the spin electric coupling (SEC) in the linear regime of FePc and Fe-FePc. Here, we perform a Taylor expansion of Eq. (2) in the main text and find a linear approximation of the exchange field:

$$B_{\text{tip}} \approx B_0 - \frac{P\gamma_T}{2\pi}\left(\ln\left(\left|\frac{\epsilon}{\epsilon+U}\right|\right) + \left(\frac{1}{\epsilon} - \frac{1}{\epsilon+U}\right) \cdot V_{\text{DC}}\right) \tag{S2}$$

To quantify the frequency shift with voltage we performed a linear fit to the peak positions $f_{res}$ over $V_{DC}$:

$$f_{\text{res}} = m \cdot V_{\text{DC}} + c \tag{S3}$$

$$c = \frac{g\mu_B}{h}\left(B_0 - \frac{P\gamma_T}{2\pi}\ln\left(\left|\frac{\epsilon}{\epsilon+U}\right|\right)\right) \tag{S4}$$

$$m = -\frac{g\mu_B}{h}\frac{P\gamma_T}{2\pi}\left(\frac{1}{\epsilon} - \frac{1}{\epsilon+U}\right) \tag{S5}$$

With the offset $c$ and slope $m$. The latter captures the magnitude of the SEC and the direction of the frequency shift via its sign. Fig. S4 a) provides an overview of the extracted $m$ over the setpoint conductance $G$ (reflecting different tip-sample distances) for various tip-molecule pairs. We find that both the magnitude and sign of $m$ vary for different tips, with extracted values ranging from $0.44$ MHz/mV to $8.22$ MHz/mV.

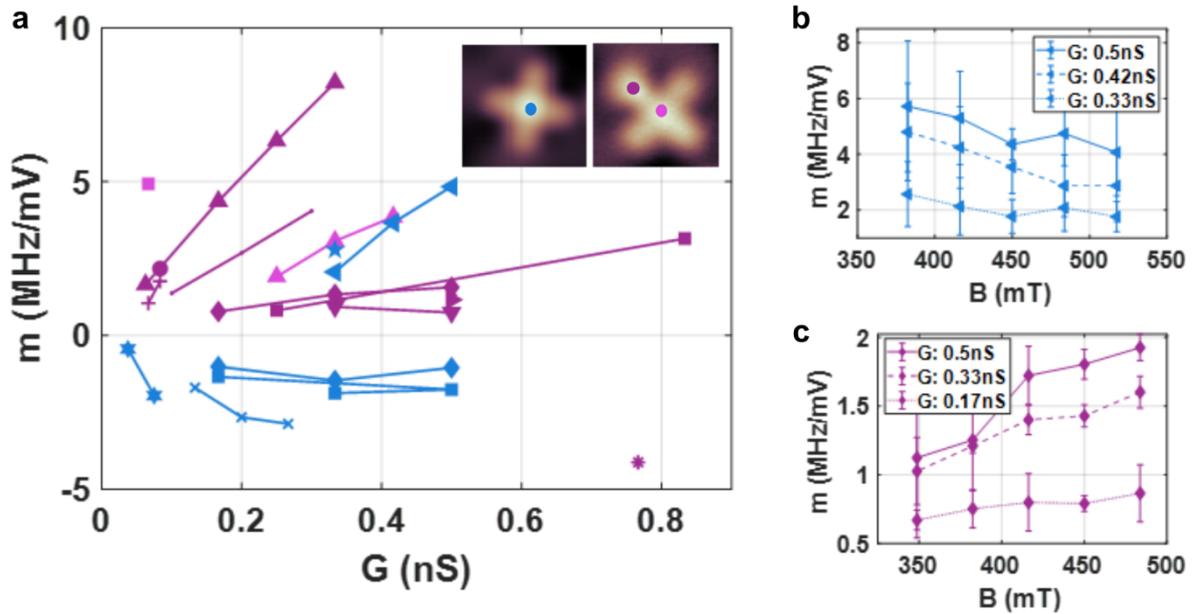

**Supplementary Fig. S4. (a)** Extracted slope $m$ of the resonance frequency ($f_{\text{res}} = m \cdot V_{\text{DC}} + c$) shift in the linear regime plotted over the setpoint conductance $G$ of the respective measurement. Different symbols correspond to different tips; connected data points represent measurements at various $G$ with the same tip. Blue: measurements on the FePc center. Pink: measurements on the Fe-FePc center. Purple: measurements on the Fe-site of the Fe-FePc. The measurement positions are also indicated in the inserted topographies of the respective molecules. **(b)** $m$ as a function of the external magnetic field $B$ for different $G$ on the FePc. **(c)** $m$ as a function of $B$ for different $G$ on the Fe-site of Fe-FePc.

In most cases, higher conductance values $G$, corresponding to smaller tip-sample distances, lead to larger SEC $|m|$. This is consistent with the expectation from a larger tip-atom coupling $\gamma_T$. However, in some cases the SEC does not increase with higher $G$, suggesting that additional mechanisms influence $f_{\text{res}}$. This could stem from the contributions of magnetic dipole-dipole interaction of the tip field[15] or from a piezoelectric effect[16]. We find a tendency for the Fe–FePc complex (center and Fe-site) to exhibit positive $m$ ($P < 0$) whereas FePc tends to show negative $m$ ($P > 0$). Additionally, in certain cases, the strength of the SEC appears to vary with the external magnetic field, as illustrated in Fig. S4b and S4c. This magnetic field dependence could arise from a field-induced change in the effective g-factor under applied bias[16].

A more trivial explanation is a dependence of the magnetic tip field on the external magnetic field $B_{\text{tip}}(B_{\text{ext}})$, i.e. a paramagnetic tip which changes its spin polarization $P$ with $B_{\text{ext}}$.

## 7. Polarity Preference of FePc and Fe-FePc

Studying the SEC, we found a strong dependence of the ESR amplitude on the bias polarity. For FePc, the ESR signal is typically stronger for positive $V_{\text{DC}}$ (Fig. S5a), whereas the Fe-FePc complex generally shows a more prominent signal for negative $V_{\text{DC}}$ (Fig. S5b or Fig. S10). This polarity-dependend trend was consistently observed for all ESR-active tips. Within the framework of the exchange bias model, this effect can be caused by an enhanced Rabi rate close to the energy levels of the SIAM[10,12]. Since the two orbitals are not equally distant from the Fermi energy, we expect an enhanced Rabi rate for the FePc (Fe-FePc) at positive (negative) $V_{\text{DC}}$. This results in an enhanced ESR amplitude for the respective voltage polarity.

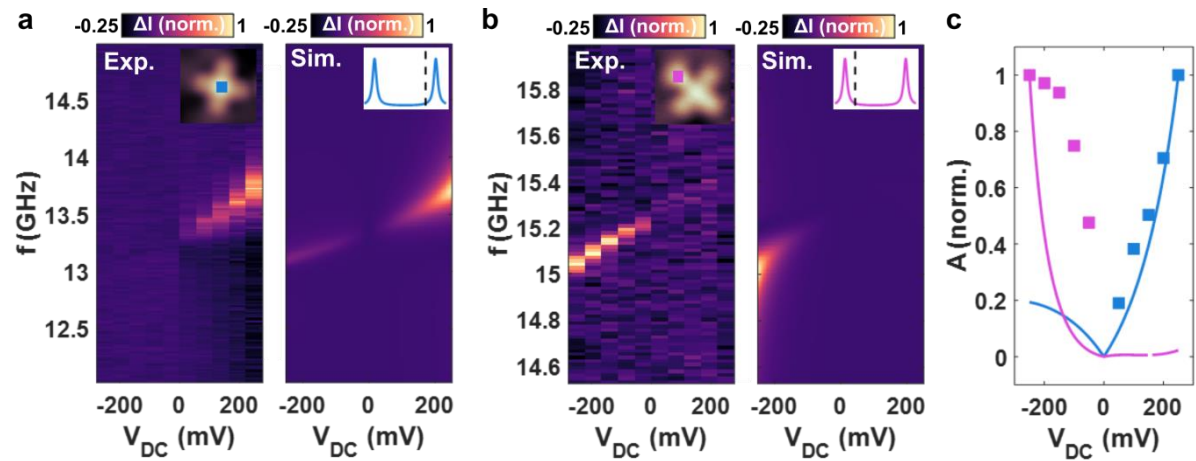

**Supplementary Fig. S5. (a)** Left: ESR frequency sweeps on FePc (see inserted topography) for different bias voltages plotting $\Delta I$ as a function of $f$ and $V_{\text{DC}}$ (ESR conditions: $I_{\text{set}} = 5$ pA, $V_{\text{set}} = 100$ mV, $B = 484$ mT, $V_{\text{RF}} = 10$ mV). Right: Corresponding simulation based on the transport model (Simulation parameters: $\epsilon = -2000$ meV, $U = 2500$ meV, $P = 0.7, \gamma_T = 15$ µeV, $\gamma_S = 10$ µeV, $T = 1$ K, $\Delta = 20$ meV, $B = 455$ mT, $\theta = 15°$, $g = 2$). Both colormaps were normalized to the maximum value. The inset on the top right shows a schematic plot of the SIAM's spectral function with the doubly occupied orbital close to the Fermi level (dotted line). **(b)** Left: Measurements on the Fe-FePc complex (see inserted topography) showing $\Delta I$ as a function of $f$ and $V_{\text{DC}}$ (ESR conditions: $I_{\text{set}} = 10$ pA, $V_{\text{set}} = -60$ mV, $B = 484$ mT, $V_{\text{RF}} = 8$ mV). Right: Corresponding simulation using the electronic properties of the Fe-FePc complex (simulation parameters: $\epsilon = -350$ meV, $U = 1000$ meV, $P = 0.7, \gamma_T = 5$ µeV, $\gamma_S = 10$ µeV, $T = 1$ K, $\Delta = 20$ meV, $B = 549$ mT, $\theta = 15°$, $g = 2$). Both colormaps were normalized to the maximum value. The inset on the top right shows a schematic plot of the SIAM's spectral function with the single unoccupied orbital close to the Fermi level. **(c)** Extracted ESR amplitudes $A$ as a function of $V_{\text{DC}}$. The values obtained from the experiment are shown as squares (blue for FePc, purple for Fe–FePc), and simulation results are plotted as solid lines.

We reproduce the two experimental datasets presented in Fig. S5 quantitatively by simulations based on the exchange bias[11,12] using the code provided by Ref.[17]. We set the chemical potential of the sample to zero. Further, we fixed the temperature, g-factor, the broadening of the orbital and the modulation of the tunneling matrix element: $T = 1\text{ K}, g = 2, \Delta = 20\text{ meV}, A_T = 50\%$. The ionization energy and Coulomb repulsion energy are chosen to match the orbital structure of FePc (Fig. S5a, right: $U = 2500\text{ meV}, \epsilon = -2000\text{ meV}$) and Fe-FePc (Fig. S5b, right: $U = 1000\text{ meV}, \epsilon = 350\text{ meV}$). To match the experiment we adapted the magnetic field $B$, the angle between the polarization axis and the external magnetic field $\theta$, the tip polarization $P$ and the coupling to the substrate (tip) $\gamma_S$ ($\gamma_T$) as stated in the description of Fig. S5a,b. Fig. S5c shows a plot of the extracted amplitudes $A(V_{DC})$ from both experiment and simulations. We find a qualitative agreement between experiment and simulations. For a more quantitative explanation, we believe that additional effects need to be considered, such as spin-pumping.

## 8. Piezo-Electric Displacement

We also attempted to explain the spin-electric coupling (SEC) observed for FePc using a piezoelectric displacement model as described in Ref.[16].

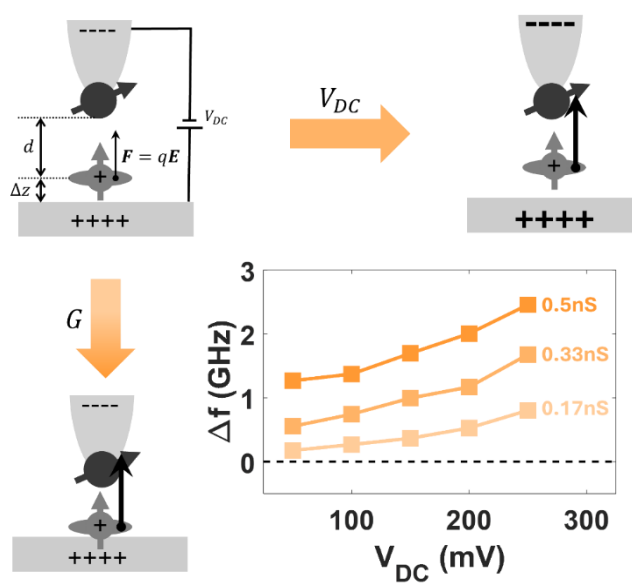

**Supplementary Fig. S6.** Sketch of the tunnel junction with the magnetic tip and the spin on the surface. The applied bias voltage $V_{DC}$ generates an electric field across the junction, illustrated by the positive and negative signs (corresponding to a positive $V_{DC}$). The partially charged surface spin is distorted by the electric force $\boldsymbol{F} = q\boldsymbol{E}$ and therefore experiences a different magnetic field from the tip. The distortion is varied either via the setpoint conductance $G$ or the applied bias voltage $V_{DC}$. The plot in the lower left exemplary shows measured data, displaying $\Delta f = f_{\text{res}} - 2\mu_B B/h$ as a function of $V_{DC}$ for three different conductance values: $0.17\text{ nS}, 0.33\text{ nS},$ and $0.5\text{ nS}$ (ESR conditions: $V_{\text{set}} = 60\text{ mV}, B = 383\text{ mT}, V_{RF} = 10\text{ mV}$).

In this model, the electric field emanating from the tip apex $|\mathbf{E}| = V_{\mathrm{DC}}/d$ leads to a force acting on the charged surface molecule, that is modelled by a displacement force $F_{\mathrm{dis}} = -k \cdot \Delta z$. In the equilibrium state $F_{\mathrm{dis}} + F_{\mathrm{el}} = 0$ and consequently $F_{\mathrm{dis}} = k \cdot \Delta z = F_{\mathrm{el}} = qE = q \cdot V_{\mathrm{DC}}/d$. Thus, $\Delta z = q \cdot V_{\mathrm{DC}}/(d \cdot k)$. Therefore, in case of $V_{\mathrm{DC}} > 0$ and a positive charge $q > 0$, the spin would move closer to the tip ($\Delta z > 0$), while it would move away from the tip for a negative charge $q < 0$ ($\Delta z < 0$). For a higher, positive displacement $\Delta z$, the on-surface spin is closer to the magnetic tip and thus experiences a stronger field. Analogously, if it is further away, the magnetic field is weaker, thus $B_{\mathrm{tip}} \propto \Delta z$. The direction of the frequency shift, then depends on whether $B_{\mathrm{tip}}$ is aligned with $B$ (positive shift) or anti-aligned (negative shift). The presented data set of Fig. S6 (same molecule/tip as in Fig. 2b in the main text) shows the shift of the resonance peak $\Delta f$ as a function of setpoint conductance $G$ and applied bias voltage $V_{\mathrm{DC}}$. We observe that $\Delta f$ is positive and increases for a decreasing tip-sample distance $d$ (i.e. higher $G$). Therefore, the spin must experience a higher magnetic tip field when the tip is moved closer to it which has the same effect as a movement $\Delta z > 0$.

Since we also observe an increase of $\Delta f$ with $V_{\mathrm{DC}}$, we also conclude a positive $\Delta z$ through the electric field as it appears for a positive charge $q > 0$. However, previous studies have assigned a negative partial charge to the FePc molecule[2]. If the molecule indeed carries a negative charge, it would be repelled by the electric field under positive bias, resulting in a frequency shift in the opposite direction. In all of our measurements with different tips and molecules the positive charge was concluded when applying a piezoelectric displacement model. Therefore, the observed SEC for FePc is unlikely dominated by a piezoelectric displacement of the surface spin in the magnetic field gradient of the tip.

Another model for the SEC is a modulation of the g-factor by the electric field[16]. While we do not exclude that the g-factor is influenced by the electric field, we observe in our experiments that the SEC varies for different tips as presented in Fig. 2 and Supplementary section 6. In contrast SEC stemming from g-factor modulation would be tip-independent.

## 9. Non-linear Spin Electric Coupling of FePc

In this section we provide further information on the non-linearity of the SEC on FePc which is observed in ESR frequency sweeps. Fig. S7 shows a selection of the data of Fig. 2a and b in the main text as raw data waterfall plots. The plots capture the drastic increase of the resonance amplitude at high $V_{\mathrm{DC}}$ together with the non-linear shift of the resonance position and in the case of Fig. S7a an increased asymmetry of the peak line shape. We also see an increased noise level and additional features stemming from the mismatches in the transfer function for high $V_{\mathrm{DC}}$ due to the large currents. However, in all cases the fit to a Fano Lorentzian still shows good agreement.

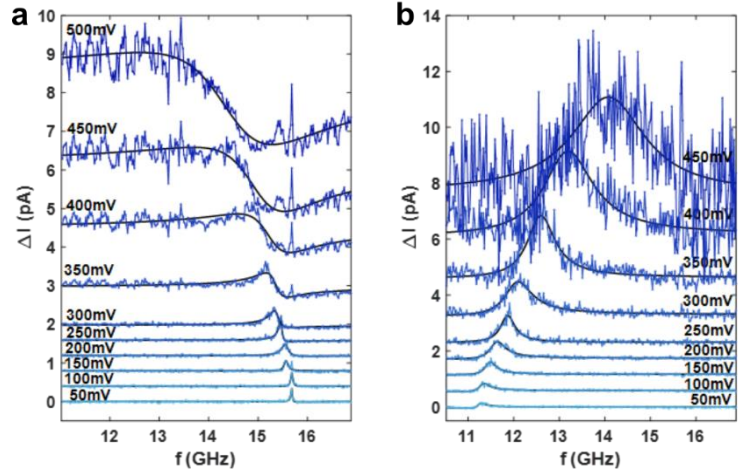

**Supplementary Fig. S7. (a)** Waterfall plot of individual frequency sweeps from the experimental data presented Fig. 2a. The overlaid black lines represent fits to the Fano function shown in Eq. (S1). **(b)** Waterfall plot of frequency sweeps from the experimental data presented in Fig. 2b with the Fano fit in black.

As mentioned in the main text we fitted the experimentally determined resonance frequencies to the exchange bias model described by Eq. (2) (see Fig. S8): As measurements were only performed at positive bias voltages, we fixed $\epsilon = -2000$ meV, consistent with the value obtained from d$I$/d$V$ measurements. To incorporate all pre- factors of $B_{\text{tip}}$ into the coupling term $\gamma_T$, we fixed the magnitude of the tip spin polarization to $|P| = 1$. Notably, $\gamma_T$ contains the angle between polarization axis and the external magnetic field $\theta$ which is used in the simulations later. Additionally, we set the g-factor to $g = 2$, attributing any deviation from the exchange bias model to the residual magnetic field $B_0$. For the dataset shown in Fig. 2a (red), we find the following fit parameters: $P = 1, \gamma_T = 100 \pm 9$ mT, $\epsilon + U = 553 \pm 6$ meV, $B_0 = 0.1 \pm 2.7$ mT. And for Fig. 2b (blue): $P = -1, \gamma_T = 225 \pm 24$ mT, $\epsilon + U = 477 \pm 10$ meV, $B_0 = -57 \pm 7$ mT. Therefore, we find a higher coupling to the tip $\gamma_T$ for the blue dataset as expected for the higher setpoint conductance used. Although, this could also be explained by a higher tip-polarization. For the blue dataset we find a quite substantial residual tip field $B_0$ which hints towards further contributions from other mechanisms e.g. magnetic dipole-dipole interaction.

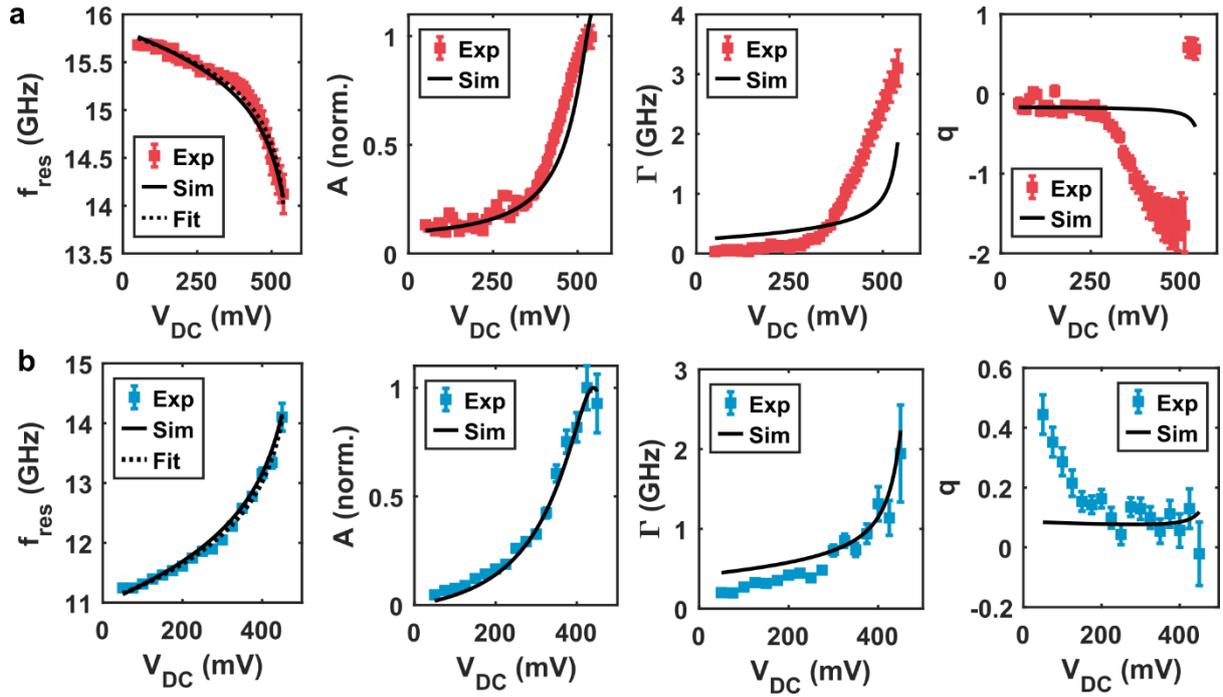

**Supplementary Fig. S8. (a)** Comparison between experiment and simulation for the data presented in Fig. 2a. The panels present the extracted parameters from Eq. (S1) ($f_{\text{res}}, A, \Gamma, q$) to the respective frequency sweeps as a function of $V_{\text{DC}}$. The resonance frequencies obtained from the experiment are shown in red, while simulation results are shown in black. The leftmost panel showing $f_{\text{res}}$ additionally includes the fit to the analytical formula presented in Eq. (2). **(b)** Same analysis as in (a), applied to the data from Fig. 2b with the experiment in blue.

For the full simulation of the ESR signal we used the obtained energy levels $\epsilon$ and $\epsilon + U$ from the previous fit to Eq. (2). We found best agreement with the experiment for the simulation parameters presented in Tab. 1 where we focused on reproducing $f_{\text{res}}(V_{\text{DC}})$. The result is shown in the panels of Fig. S8. We find a very good agreement between experiment and the model for the resonance frequency $f_0$ and amplitude $A$. Further, the simulation qualitatively captures the observed trend in linewidth $\Gamma$ of the resonance peaks. However, quantitative discrepancies remain, which are most prominent in the red dataset. Additionally, the evolution of the Fano asymmetry parameter $q$ is not well described by the model. The discrepancies found can be explained primarily by the increase in conductance as $V_{\text{DC}}$ approaches the prominent peak obtained in the d$I$/d$V$ measurements: The resulting non-linear change of the tunnel current leads to a rapid increase of $A$ and $\Gamma$ and a change of $q$. This conductance change is not included in the simulation. Further contributions such as spin-pumping are also possible. Despite these limitations, the comparison proves the ability of the exchange bias model in describing the spin electric coupling of single spins on surfaces.

|       | $B$     | $\theta$ | $P$   | $\gamma_S$ | $\gamma_T$ | $U$       |
|-------|---------|----------|-------|------------|------------|-----------|
| Red   | 589 mT  | 17.5°    | −0.66 | 17 μeV     | 19 μeV     | 2553 meV  |
| Blue  | 337 mT  | 14.5°    | 0.72  | 32 μeV     | 36 μeV     | 2477 meV  |

**Supplementary Table 1.** Simulation parameters of the colormaps presented in Fig. 2. With the magnetic Field $B$, the angle between the polarization axis and the external magnetic field $\theta$ (see Ref.[18]), the tip polarization $P$, the coupling to the substrate (tip) $\gamma_S$ ($\gamma_T$) and the already introduced Coulomb repulsion energy $U$. For both simulations we fixed: $T = 1\,\text{K}, g = 2, \Delta = 20\,\text{meV}, A_T = 50\%, \epsilon = 2000\,\text{meV}$.

Besides the demonstrated dependence on the tip-polarization $P$, the magnitude of the exchange bias also depends on the coupling between tip and molecule $\gamma_T$, which can be altered by varying the thickness of the vacuum tunnel barrier. This is achieved by varying the conductance setpoint $G$ with the relation $\gamma_T \propto G$. As shown in Fig. S9, the magnitude of the SEC increases for positive and negative polarizations with increasing $G$.

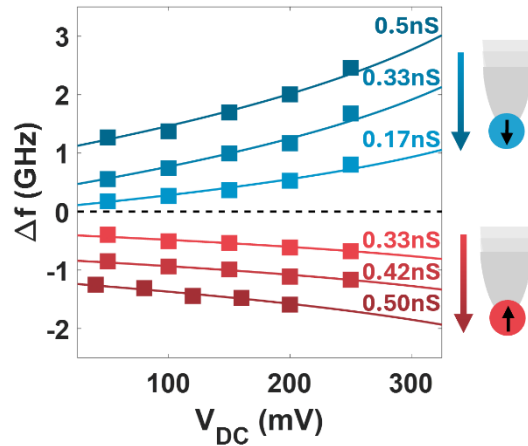

**Supplementary Fig. S9.** Frequency shift $\Delta f(V_{DC})$ for different setpoint conductance $G$ (ESR conditions, red: $V_{set} = 60\,\text{mV}, B = 500\,\text{mT}, V_{RF} = 10\,\text{mV}$; blue: $V_{set} = 60\,\text{mV}, B = 383\,\text{mT}, V_{RF} = 10\,\text{mV}$). The solid lines show fits to the data according to the exchange bias model. The measurements allow the assignment of the direction of $B_{tip}$: The magnitude of $\Delta f$ increases with G since $B_{tip} \propto G$ in good approximation[15,19] and the sign of $B_{tip}$ can according to Eq. (1) be interpreted as opposite tip polarities (see sketches).

In Fig. S10 we show the magnetic field dependence of the resonance peak in the non-linear regime for three magnetic fields: $450\,\text{mT}, 518\,\text{mT}, 585\,\text{mT}$. The color plots in Fig. S10a and the extracted resonance frequencies in Fig. S10b show a clear shift of the feature with magnetic field attributing it to the spin resonance of the surface spin. The data was obtained on the same molecule as Fig. 2a but with a different tip.

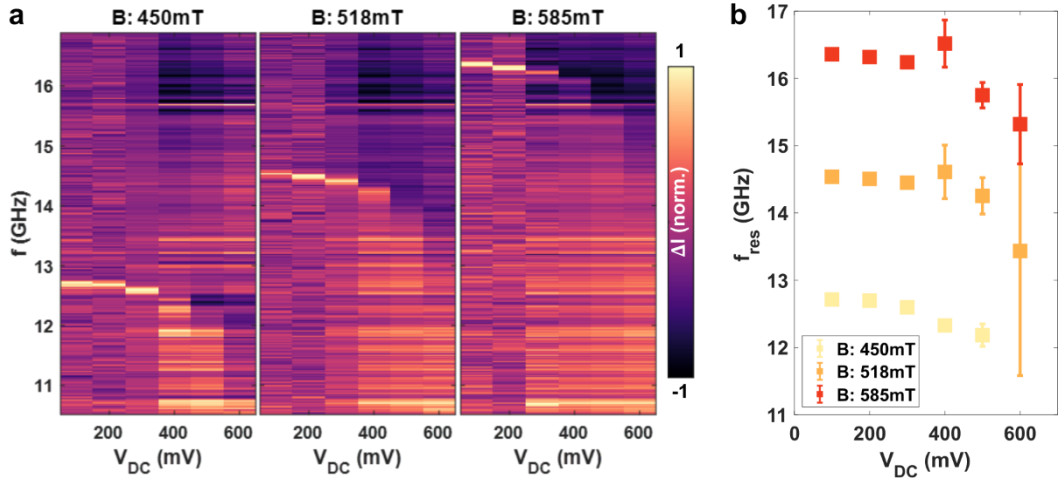

**Supplementary Fig. S10. (a)** Colormaps of $\Delta I$ as a function of $V_{DC}$ and $f$ on FePc for different external magnetic fields $B$ (ESR conditions: $I_{set} = 4$ pA, $V_{set} = 60$ mV, $V_{RF} = 12$ mV). For better visibility each frequency sweep is normalized separately. **(b)** Extracted resonance frequencies $f_{res}$ from (a) as a function of $V_{DC}$.

## 10. Non-linear spin electric coupling of Fe-FePc

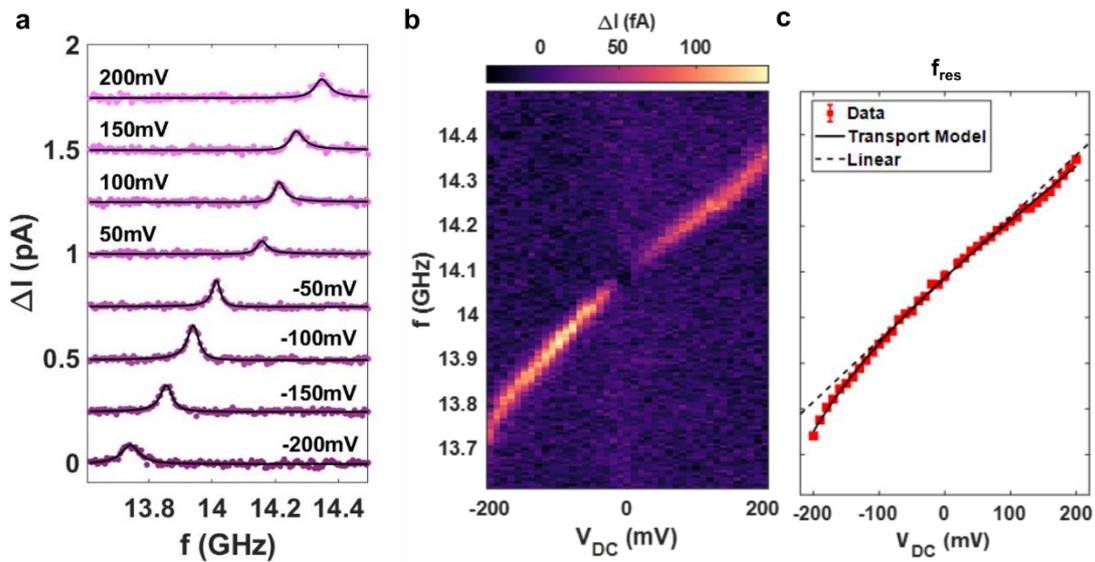

**Supplementary Fig. S11. (a)** Waterfall plot of frequency sweeps showing $\Delta I$ as a function of $f$ for several $V_{DC}$ measured on the Fe-site of a Fe-FePc complex (ESR conditions: $I_{set} = 6$ pA, $V_{set} = -60$ mV, $V_{RF} = 10$ mV, $B = 469$ mT). The solid black lines represent Fano fits to the experimental data (purple). With changing $V_{DC}$ we see a clear shift of the resonance position. **(b)** Corresponding colormap $\Delta I(f, V_{DC})$ of the data set, revealing a higher amplitude for negative voltages. **(c)** Extracted resonance frequency $f_0$ over $V_{DC}$. A linear fit (dashed line) ranging from $-100$ mV to $100$ mV deviates at higher voltages from the experimental data. A fit using the exchange bias model (solid line) described by Eq. (2) in the main text captures the non-linear behavior at higher voltage magnitudes. The fitted parameters are: $\epsilon = -344 \pm 47$ meV, $U = 998 \pm 269$ meV, $B_0 = 41 \pm 4$ mT, $\gamma_T = 66.5 \pm 14.8$ mT, $P = -1$. We find $\epsilon$ and $U$ to be consistent with the values obtained from d$I$/d$V$ measurements (see Fig. S1).

## 11. Rabi Detuning Measurements

Fig. S12a shows a waterfall plot of the Rabi oscillations which were detuned by changing the frequency $\delta f = f - f_{\text{res}}$, shown in Fig. 3c. Each trace was fitted according to Eq. (3) in the main text. As can be seen from Fig. S12b,c the extracted Rabi rate $\Omega$ and amplitude $A$ match the expected behaviors from Eq. (4) in the main text. In Fig. S12d we use the fitted datasets in b,c to simulate the observed experimental data, showing good agreement. $T_2$ and $\phi$ were chosen according to the mean values obtained from the experiment.

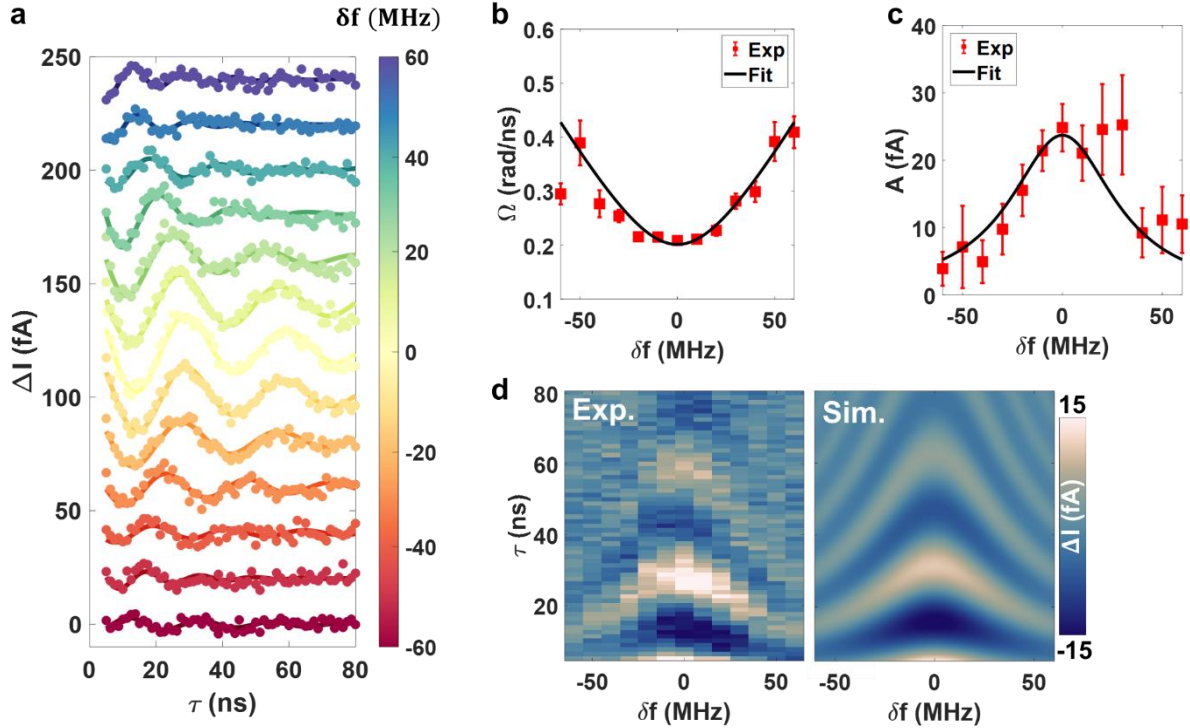

**Supplementary Fig. S12.** (a) Waterfall plot of Rabi oscillation measurements, corresponding to Fig. 3c in the main text, recorded at various frequency detuning $\delta f$. The traces showing $\Delta I$ over $\tau$ are vertically shifted for clarity. Experimental data points are shown as circles, and the solid lines represent fits to Eq. (3). (b) Extracted Rabi rates $\Omega$ as a function of $\delta f$, obtained from the fits in panel (a). The solid curve is a fit to Eq. (4) in the main text, yielding an intrinsic Rabi rate of $\Omega_0 = 0.201 \pm 0.006$ rad/ns. (c) Extracted amplitudes $A$ from panel (a) with a fit to Eq. (4), yielding $A_0 = 23.7 \pm 1.6$ fA. (d) Left: Colormap of the measurements from (a) showing $\Delta I$ as a function of $\tau$ and $\delta f$. Right: Simulated chevron pattern using Eq. (3) and (4) with $A_0 = 23.7$ fA, $\Omega_0 = 0.2$ rad/ns, $T_2 = 27.6$ ns, $\phi = 12.9°$.

Fig. S13a shows the same analysis for the voltage-detuned Rabi oscillations displayed in Fig. 3d in the main text (displayed again in Fig. S13d). We observe again a chevron pattern, demonstrating the detuning through the SEC. Further, the pattern appears slightly distorted. This is reflected in the extracted Rabi rates $\Omega$ and amplitudes $A$ (see Fig. S13b,c). We explain this feature via an additional modulation of $\Omega_0$ with $V_{\text{DC}}$, which is predicted by the exchange bias model (see Ref.[10,12]). In first approximation, we consider a linear contribution:

$$\Omega = \sqrt{\Omega_0^2 + \alpha^2(V_{\text{DC}} - V_{\text{set}})^2}, \quad \Omega_0 = \Omega_{00} + \beta(V_{\text{DC}} - V_{\text{set}}) \quad \text{(S6)}$$

Here, $\alpha$ is parameterizing the frequency shift with voltage detuning $\delta V = (V_{\text{DC}} - V_{\text{set}})$ from the setpoint voltage. $\Omega_{00}$ is the intrinsic Rabi rate at $V_{\text{set}}$ and $\beta$ the linear factor accounting for the additional dependence of $\Omega_0$ on $V_{\text{DC}}$. By fitting the combined formula to the experimental data in Fig. S13b we find the best agreement for $\Omega_{00} = 0.34 \pm 0.02\ \text{rad/ns}, \alpha = 2.1 \pm 0.4\ \text{MHz/mV}, \beta = -0.6 \pm 0.2\ \text{MHz/mV}$. Comparing $\alpha$ to the extracted magnitudes of the SEC $m$ from section S6 we find that it nicely fits into the observed range. Also, $\beta$ lies in the same order of magnitude. The negative sign means a stronger increase of $\Omega$ with negative $\delta V$ compared to positive $\delta V$ which as it is observed in the extracted data. This is expected from the $\epsilon$ and $U$ values extracted for the Fe-FePc complex, since we have the HOMO closer to the Fermi level as the LUMO and therefore expect a stronger virtual tunneling for negative $V_{\text{DC}}$.

For the amplitude $A$ we further assume a scaling with the tunnel current in the junction. We result with the following expression:

$$A = A_0 \frac{\Omega_0^2}{\Omega^2} \cdot \left|\frac{V_{\text{DC}}}{V_{\text{set}}}\right| \quad \text{(S7)}$$

With the amplitude at the setpoint voltage $A_0$. A fit to the experimental data is shown in Fig. S13c finding $A_0 = 21.8 \pm 3.3\ \text{fA}$. Generally, we find a good description of the experiment by this formula with some deviations for $\delta V = -20\ \text{mV}$ and $-25\ \text{mV}$ which most likely stems from a bad fit to the strongly suppressed Rabi oscillations. In Fig. S13b,c we additionally plot fits without the additional correction, i.e. $\beta = 0$. While the agreement is still reasonable, it is not as good as in the case of frequency sweep (Fig. S13) or with the inclusion of a change in $\Omega \propto V_{\text{DC}}$. We interpret these results as evidence that $\Omega_0$ is tuned via exchange bias, in line with theoretical expectations[12]. However, due to the subtle nature of the changes observed, we emphasize that further investigation is required to confirm this mechanism. Considering all dependencies we can reproduce the chevron pattern as demonstrated in Fig. S13d showing a similar tilt as the experimental data. Overall, we find a quantitative understanding of frequency and voltage detuning in Rabi measurements. Additionally, we find experimental fingerprints of the SEC stemming from the exchange bias model.

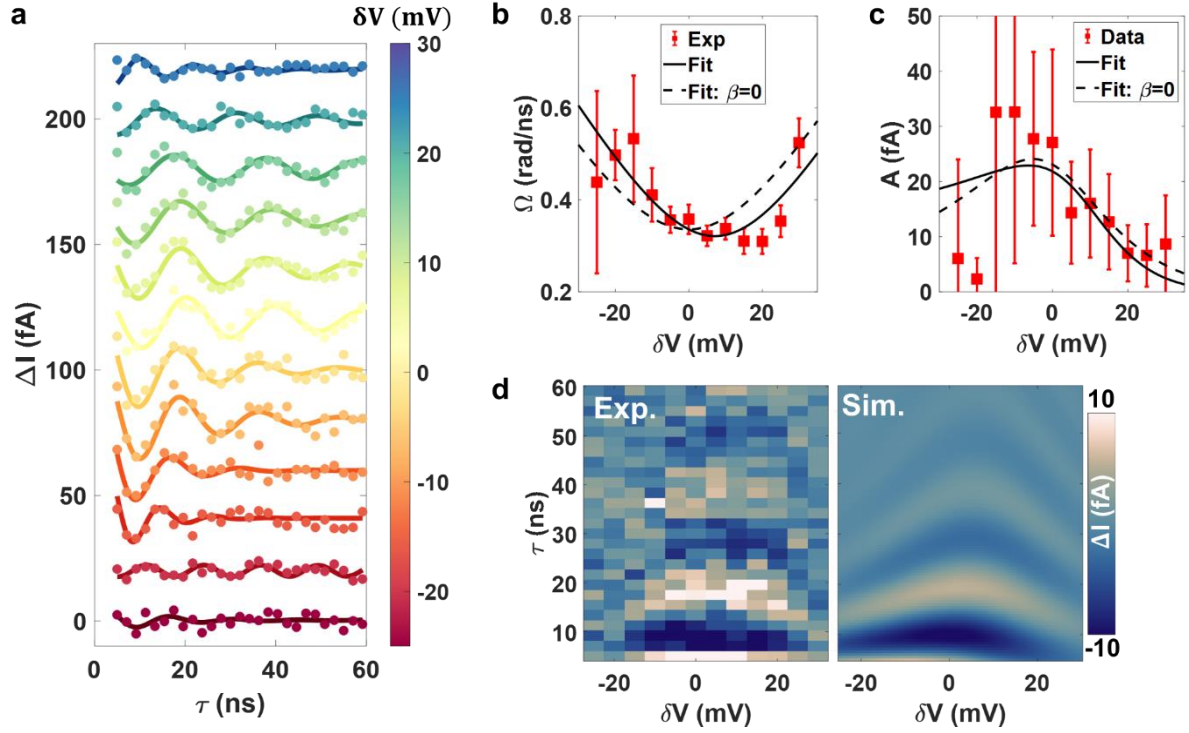

**Supplementary Fig. S13.** **(a)** Waterfall plot of the Rabi measurements from Fig. 3c measured at different voltage detuning $\delta V$. The solid lines in slightly darker colors as the round datapoints from the experiment show fits to Eq. (3). **(b)** Extracted Rabi rates from the fits of (a) as a function of $\delta V$. The solid line shows a fit to Eq. (S6). **(c)** Extracted amplitudes from the fits of (a) as a function of $\delta V$. The solid line shows a fit to Eq. (S7). **(d)** Left: Colormap of the measurements from (a) showing $\Delta I$ as a function of $\tau$ and $\delta V$. **Right:** Simulated chevron pattern using Eq. (S6) and (S7).

## 12. SEC in a coupled spin system

The coupled spin system shown in Fig. 4 in the main text can be described by two coupled spin ½ as already demonstrated by earlier works[2,20–22]. Here, the eigenstates of the system can be tuned between Zeeman eigenstates and mixed eigenstates depending on the coupling strength and the Zeeman energy difference $\delta E = g_1 \mu_B (B + B_{\text{tip}}) - g_2 \mu_B B$ of the two spins. For the measurements in Fig. 4 the system was operated in the Zeeman state regime ($G = 0.2$ nS), where the eigenstates are product states of the Zeeman states describing each spin individually. This leads to two resonance peaks in the spectrum. The system can be tuned via the tip magnetic field $B_{\text{tip}}$ to a regime where the Zeeman energies of both spins are equal ($\delta E = 0$) and we get mixed states as eigenstates. The mixed states allow for addressing four transitions. Thus, we see four peaks in the frequency sweep at the avoided level-crossing ($G = 0.6$ nS). As shown in Fig. S14a the detuning is typically achieved by changing $B_{\text{tip}}$ via the setpoint conductance $G$.

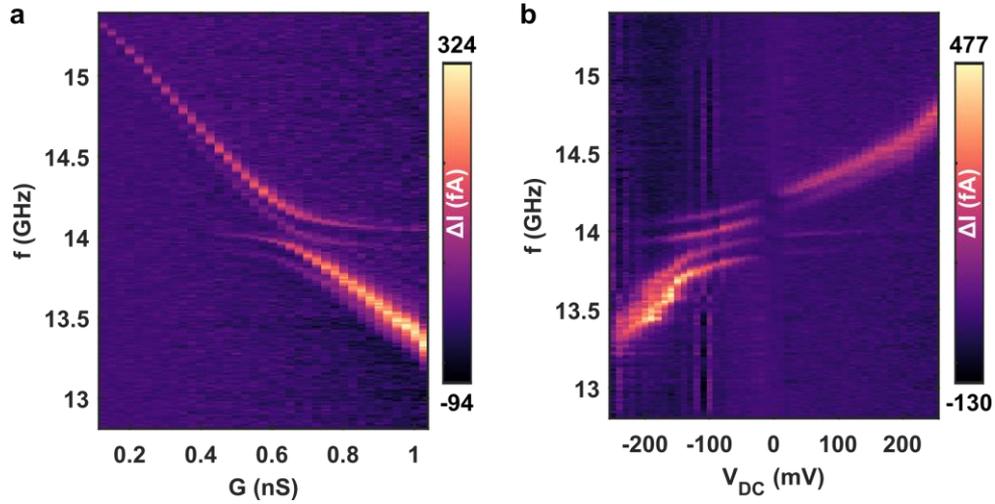

**Supplementary Fig. S14. (a)** Colormap of $\Delta I$ as a function of $f$ and setpoint conductance $G$ of the coupled spin system from Fig. 4 (ESR conditions: $V_{\text{set}} = 40$ mV, $V_{\text{RF}} = 10$ mV, $B = 494$ mT) **(b)** Colormap of $\Delta I$ as a function of $f$ and $V_{\text{DC}}$ (ESR conditions: $I_{\text{set}} = 20$ pA, $V_{\text{set}} = -60$ mV, $V_{\text{RF}} = 10$ mV, $B = 494$ mT).

Given the SEC we can use $V_{\text{DC}}$ to obtain the same behavior of the resonance frequencies (Fig. S14b). Therefore, the SEC provides a fast alternative to tune the eigenstates of a coupled spin system without any movement of the STM tip. Here, our observations match the eigenstate tuning which was reported for single Ti atoms[16]. Notably, this possible tuning of the eigenstates requires the SEC to be localized to the first spin.

In the following section we further evaluate the localization of the SEC. We can use the setpoint dependent measurements to determine the resonance frequency of the first and second spin simultaneously while keeping the tip only above the first spin: For the first spin, it can be extracted directly from the frequency sweeps. It follows a linear dependency on the setpoint conductance $G$ as shown in Fig. S14a. As described above, the Zeeman energies of the two spins are equal at the avoided level crossing. Thus, we can read of the resonance frequency of the second spin directly from this position as marked by the dotted white lines in Fig. S15a. To investigate the SEC, we evaluated this for different bias voltages: $-60$ mV, $-40$ mV, $-30$ mV, $30$ mV, $40$ mV, $60$ mV. Fig. S15a shows two examples. Fig. S15b shows the combined evaluation where the frequency shift $\Delta f = f_{\text{res}} - 2\mu_B B/h$ is plotted over $V_{\text{DC}}$. While the frequency shift of the first spin (red) changes with $V_{\text{DC}}$ and $G$ according to the SEC, the frequency shift of the second spin (blue) remains constant. Therefore, the reported SEC only acts locally on the spin underneath the tip. This localization is in-line with the mechanism of the exchange bias model and is an important demonstration of nanometer scale electric control.

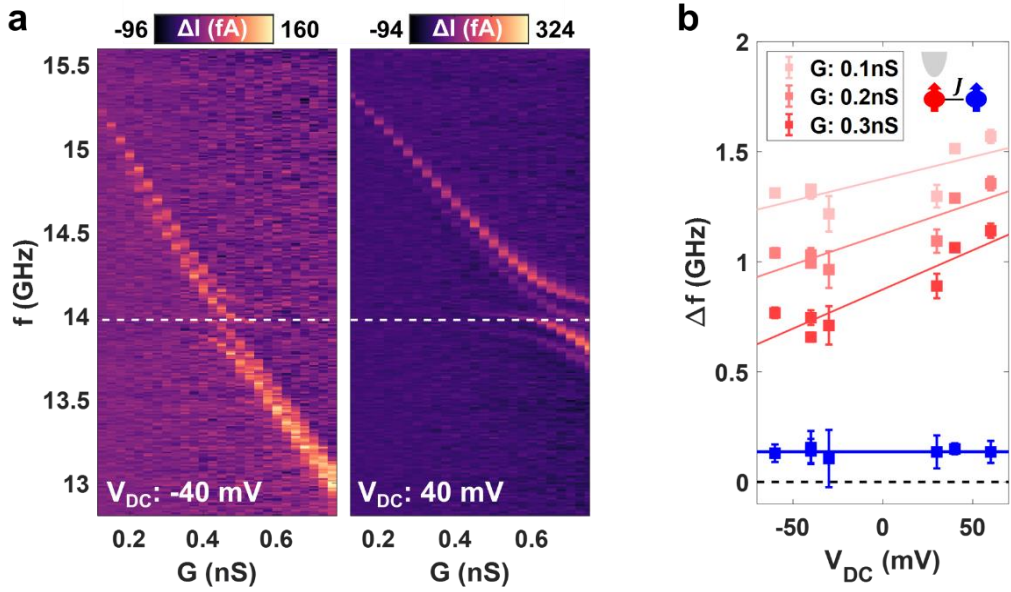

**Supplementary Fig. S15.** (a) Colormaps of $\Delta I$ as a function of $f$ and $G$ showing the avoided level crossing of the coupled spin system at $V_{DC} = -40$ mV and $V_{DC} = 40$ mV (ESR conditions: $V_{RF} = 10$ mV, $B = 494$ mT). The white dashed lines show the frequency position of the avoided level crossing (b) Extracted frequency shift $\Delta f = f_{res} - 2\mu_B B/h$ as a function of $V_{DC}$. Blue: Second spin with the mean value illustrated by the straight line. Red: Shift of the first spin for different $G$ with linear fits shown as solid lines as visual aids. The inserted sketch at the top right corner reflects the measurement setup.

## 13. Rabi measurements in a coupled spin system

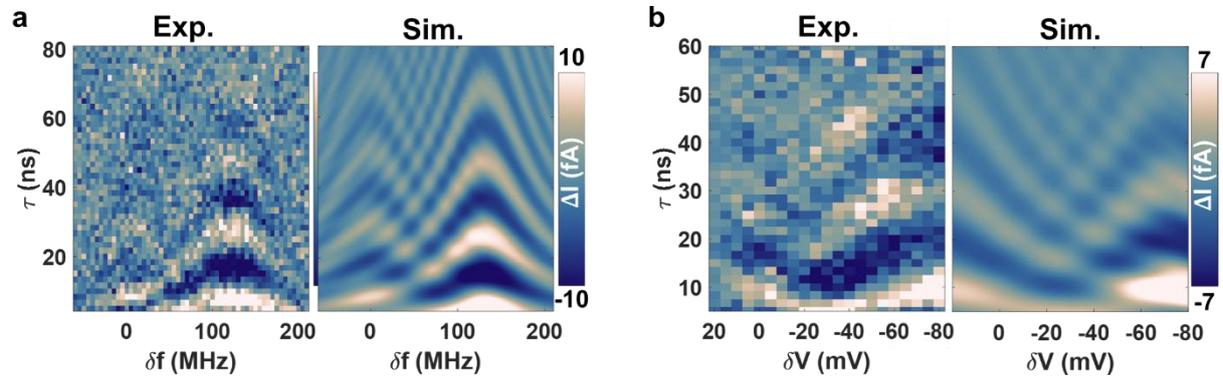

**Supplementary Fig. S16.** (a) Left: Colormap of Rabi measurements from Fig. 4 e plotting $\Delta I$ as a function of $\tau$ and $\delta f$ showing a chevron pattern with two arcs. Right: Corresponding simulation using a sum of two chevron patterns with Eq. (3-4) and: $A_1 = 8$ fA, $\Omega_1 = 0.3 \frac{\text{rad}}{\text{ns}}$, $\Phi_1 = -150°$, $T_2^1 = 30$ ns, $A_2 = 20$ fA, $\Omega_2 = 0.3 \frac{\text{rad}}{\text{ns}}$, $\Phi_2 = -100°$, $T_2^1 = 40$ ns. (b) Left: Colormap of Rabi measurements from Fig. 4 f plotting $\Delta I$ as a function of $\tau$ and $\delta V$ showing a distorted chevron pattern with two arcs. Right: Corresponding simulation using a sum of two chevron patterns created via Eq. (S6-S7) with two different amplitudes and a voltage dependent decoherence time: $V_{set}^1 = -20$ mV, $V_{set}^2 = -120$ mV, $A_1 = 8$ fA, $A_2 = 20$ fA, $\Omega_{00} = 0.3 \frac{\text{rad}}{\text{ns}}$, $\alpha = 1.5 \frac{\text{MHz}}{\text{mV}}$, $\beta = -0.4 \frac{\text{MHz}}{\text{mV}}$, $\Phi = 190°$, $T_2 = 0.25$ ns/mV $\cdot V_{DC} + 45$ ns.